# A Decade of In-text Citation Analysis based on Natural Language Processing and Machine Learning Techniques: An overview of empirical studies

Sehrish Iqbal,[a] Saeed-Ul Hassan,[a*] Naif Radi Aljohani,[b] Salem Alelyani,[c, d] Raheel Nawaz,[e] Lutz Bornmann[f]

[a] Department of Computer Science, Information Technology University, 346-B, Ferozepur Road, Lahore, Pakistan; sehrishiqbal@itu.edu.pk, saeed-ul-hassan@itu.edu.pk

[b] Faculty of Computing and Information Technology, King Abdulaziz University, Jeddah, Kingdom of Saudi Arabia; nraljohani@kau.edu.sa

[c] Center for Artificial Intelligence (CAI), King Khalid University, PO Box 9004, Abha 61413, Saudi Arabia; s.alelyani@kku.edu.sa

[d] College of Computer Science, King Khalid University, PO Box 9004, Abha 61413, Saudi Arabia; s.alelyani@kku.edu.sa

[e] Department of Operations, Technology, Events and Hospitality Management, Manchester Metropolitan University, United Kingdom; R.Nawaz@mmu.ac.uk

[f] Administrative Headquarters of the Max Planck Society, Division for Science and Innovation Studies, Hofgartenstraße, 8, 80539 Munich, Germany; lutz.bornmann@gv.mpg.de

**\*Corresponding author's email:** saeed-ul-hassan@itu.edu.pk

**Abstract** Citation analysis is one of the most frequently used methods in research evaluation. We are seeing significant growth in citation analysis through bibliometric metadata, primarily due to the availability of citation databases such as the Web of Science, Scopus, Google Scholar, Microsoft Academic, and Dimensions. Due to better access to full-text publication corpora in recent years, information scientists have gone far beyond traditional bibliometrics by tapping into advancements in full-text data processing techniques to measure the impact of scientific publications in contextual terms. This has led to technical developments in citation context and content analysis, citation classifications, citation sentiment analysis, citation summarisation, and citation-based recommendation. This article aims to narratively review the studies on these developments. Its primary focus is on publications that have used natural language processing and machine learning techniques to analyse citations.

**Keywords:** Bibliometrics, Citation Context Analysis, Citation Content Analysis, Citation Classification, Citation Sentiment Analysis, Summarisation, Recommendation

# 1    Introduction

While writing a publication, authors usually cite the publications that have influenced the research that they describe in order to explain the framework in which the research took place (Shadish et al., 1995; Turney, 2002). In the past decades, the most significant indicator of the scientific impact of a publication has been considered to be its citation count, and this has been frequently used to evaluate the performance of faculty members, research institutions, and universities (Safer and Tang, 2009). According to Anderson (2006), citation counts are not only an impact metric but also serve as an indicator to capture the overall quality of research. Citations also play a significant role by investigating both the historical roots and the novelty of new research. Through citations, measurable links can be established between the citing and cited documents to establish the 'web' of knowledge (Judge et al., 2007). Mercer, Di Marco, and Kroon (2004) proposed that the relationship between the mutually relevant publications is based on citations, while Small (2004) argues that it represents the metaphorical payment of the scholarly debt.

According to Aljaber et al (2010), the text that is used to describe the cited publication in the citing publication is termed as *citation context*; the *citation content* refers to the cited publication. In general, the citation context and citation content analysis is an extended form of the citation analysis used by the scholars to comprehend the impact of a citation. Zhang, Ding, and Milojević (2013) suggest defining citation context/content analysis as a technique to complement classic citation analyses. These context/content analyses are motivated by the need for more accurate bibliometric measures that evaluate the impact of research both qualitatively and quantitatively (Chang, 2013; Galgani et al., 2015). Applying these analyses, various schemes have been proposed to classify the functions of and the reasons for a citation – the analysis of the specific relationship between the cited and citing publications has also been conducted (Hooten, 1991; Teufel et al., 2006).

Conventional citation analysis is quantitative in nature and takes no account of contextual information, whereas citation content and context analyses consider both qualitative and quantitative factors (Cronin, 1984). In recent times, open access to full-text research publications and technical advancements have brought about extensive changes in the methods and techniques to analyse the context/content of citations (McCain and Turner, 1989). Therefore, identifying and describing these changes are an important motivation to write this review. Citation context and citation content analysis analyse the text around the citation anchors. According to Zhu, Turney, Lemire, and Vellino (2015: 9) "when a reference is mentioned in the body of the citing paper, the text that appears near the mention is called the citation context". The context is established through the citation's location in the citing text, the words around it and its semantic context, each with implications (Bornmann et al., 2018). Tahamtan and Bornmann (2019) explain the difference between context and content analyses as follows:

> "In citation content analysis, the semantic content of the text surrounding a given citation (cited document) within the citing document(s) is read to characterize the cited document. However, in citation context studies, the citing text around the reference anchor is analysed. In other words, the text around citations (cited documents) in the citing document is used to characterize citations in the citing document. It is not the objective of citation context studies to yield information about the content of a certain cited document, but to characterize the citation process of the citing authors." (p. 1652)

Over the past half-century, several reviews have been published on the topic of citation analysis. Small (1982) and Cronin (1984) presented the earliest reviews on the purpose of citation and its possible classification. A survey by MacRoberts and MacRoberts (1989) reviewed the problems in citation analysis pertaining to cursory attention, i.e., biased citations, self-citations, etc. Liu (1993) presented a review focusing on citation motivation, function, concept, and quality to explore the norms relevant in the citation process and the complexities involved in following the norms. Ding et al. (2014) conducted a survey on citation content analysis that examined the foundation, methodologies, and application. Hernández and Gómez (2015) reviewed sentiment analyses of citations, summarised the general concepts of polarity classification with the purpose of identifying trends, and suggested possible future research directions. The recent review of studies on citing processes by Tahamtan and Bornmann (2019) deals with articles published from 2005 to 2018 and focuses on authors' motivation to cite an article. The review outlines various approaches to citation classification and presents how empirical studies have investigated them.

Table 1 compares some seminal reviews of in-text citation analysis published 2004 or later, emphasising their approach and review focus. These reviews have usually focused on conventional challenges, such as citation behaviour, the role of citations, and citation classification. None have focused, however, on approaches using Machine Learning (ML) and Natural Language Processing (NLP) in citation context analyses (Yin et al., 2011), citation content analyses (Ding et al., 2013; Jeong et al., 2014), classification of citations (Cohen et al., 2006), citation sentiment analysis (Hernández and Gómez, 2015; Yousif et al., 2019), and citation summarisation (Gambhir and Gupta, 2017; Karimi et al., 2018).

**Table 1.** Previous surveys on citation analysis

| Article | Approach | Review focus |
|---|---|---|
| White (2004) | non-systematic | Reviewed the contributions in the field of information science, related to citation classification, citation content, and context analysis that were published after 1970. |
| Bornmann and Daniel (2008) | non-systematic | The authors presented a review of publications on citing behaviour by studying the research articles published from 1960 to 2005. |
| Ding et al. (2014) | non-systematic | A survey on citation content analysis has been conducted that examined the foundations, methodologies, and applications of citation content analysis. |
| Hernández-Alvarez and Gómez (2015, 2016) | non-systematic | The authors provide a survey on articles that worked on the problem of citation context identification and classification. They analysed the latest techniques and data repositories used for citation context analysis. |
| Mäntylä et al. (2018) | non-systematic | The review focuses on the top-cited papers from Scopus and Google Scholar on citation sentiment analysis that were published between 2005 and 2016. |
| Tahamtan and Bornmann (2019) | systematic | The review focuses on studies investigating the relationship between cited and citing documents. The authors review studies published between 2006 and 2018. |
| Ma et al. (2020) | non-systematic | The purpose of this review is to identify the methods and information used for citation-based recommendation systems. |

In contrast to the existing studies on citation analysis, the current review adopts a systematic approach for the collection and analysis of the studies on citation context and content analysis from Jan 2005 to Dec 2019. The review focusses on the latest technological developments from the past decade, which leverages state-of-the-art NLP and ML approaches to convert unstructured citation contexts into a useable format to obtain insights from the citation anchor data.

In recent years, due to the availability of full-text publications and improved ability to process large textual datasets, significant advances have been made in the analysis of scholarly documents (Safder and Hassan, 2019). The advances have allowed scholars to use a variety of features and ML techniques to determine the citation function, analyse citation polarities, generate citation-based summaries, and build citation-based retrieval systems. Our review comprehensively demonstrates the progress in NLP and ML techniques in this domain and investigates the methodological strengths and weaknesses of the various studies.

The prominent techniques in NLP include n-grams, bag-of-words, and word2vec, while the best-known ML classifiers are:

- Support Vector Machine (SVM: a discriminative classifier that is trained on a labelled dataset, outputting an optimal hyperplane to classify unlabelled data),
- Naïve Bayes (NB: a probabilistic classifier that refers to the conditional independence of each feature and is particularly used when the dimensionality of the inputs is high),
- Multinomial Naïve Bayes (MNB: an instance of a NB that uses a multinomial distribution for each feature),
- Hidden Naïve Bayes (HNB: an extended form of NB that retains its efficiency and simplicity while relaxing its independence assumption),
- Maximum Entropy (MaxEnt: a probabilistic classifier that finds weights for the features that maximise the likelihood of the training data),
- Decision Tree (DT: a predictive modelling approach in which trees are generated using an algorithmic technique that identifies ways to categorise a dataset, based on several conditions),
- Random Forest (RF: a non-parametric algorithm that uses multiple decision trees),
- K Nearest Neighbour (KNN: a non-parametric algorithm that classifies a data point based on what group is nearest to the particular data point), and
- Logistic Regression (LR: a type of statistical analysis that predicts the outcome of a dependent variable based on independent variables).

Some studies have used deep learning classifiers such as Artificial Neural Network (ANN: inspired by biological neural networks, i.e., the human brain, and built to simulate humans' interconnected processes), Convolutional Neural Network (CNN: a special type of neural networks designed for cognitive tasks like image processing and NLP), Recurrent Neural Network (RNN: an improved variation of neural networks with a short-term memory to retain the contextual information from earlier results), and Long Short-Term Memory (LSTM: a variant of an RNN that uses the short-term memory of RNN neurons and makes them last longer).

Table 2 shows an overview of the studies included in this review and outlining the pros and cons of the various NLP and ML techniques.

**Table 2.** Techniques used for citation content and context analysis

| Methods for citation context/content analysis | | |
|---|---|---|
| **Method name** | **Pros** | **Cons** |
| **Supervised techniques** | | |
| RF (Valenzuela et al., 2015a) | <ul><li>Efficient on large datasets</li><li>Quick to train and requires almost no input preparation</li></ul> | <ul><li>Takes up hundreds of megabytes of memory</li><li>Performs badly on time-series data</li></ul> |
| MaxEnt (Jebari et al., 2018) | <ul><li>Requires less processing</li><li>Preserves information on given data</li></ul> | <ul><li>High computational complexity</li><li>Low efficiency</li></ul> |
| LR (Small, 2018a) | <ul><li>Performs well when the dataset is linearly separable</li><li>Training is fast</li></ul> | <ul><li>Fails if decision boundary is non-linear</li><li>Overfit in high-dimensional datasets</li></ul> |
| NB (Bakhti et al., 2018) | <ul><li>Requires less training data and fast to classify</li><li>Not sensitive to irrelevant features</li></ul> | <ul><li>Assumes independence of features</li><li>Prediction accuracy is lower than that of other models</li></ul> |
| SVM (M. Wang et al., 2019) | <ul><li>Works well with a clear margin of separation</li><li>Effective in high dimensional spaces</li></ul> | <ul><li>Performs badly when the dataset is large and has more noise</li><li>Required training time is high</li></ul> |
| DT (Tuarob et al., 2019) | <ul><li>Deals with noisy or incomplete data</li><li>The ability to select the most discriminatory features</li></ul> | <ul><li>Prone to overfitting</li><li>More complex and takes high time to train</li></ul> |
| KNN (Zafar et al., 2019) | <ul><li>No training is needed</li><li>New data can be added seamlessly without impacting the accuracy</li></ul> | <ul><li>Does not work with large and high-dimension datasets</li><li>Sensitive to noisy data</li></ul> |
| **Unsupervised techniques** | | |
| Clustering (Balabantaray et al., 2015) | <ul><li>Computationally faster where variables are huge</li><li>Produces tighter clusters</li></ul> | <ul><li>Works badly with global cluster</li><li>Works badly with clusters of different density and size</li></ul> |
| SOM (Hassan, Iqbal, et al., 2018) | <ul><li>Easy to observe similarities in the data due to dimensionality reduction</li><li>Capable of handling several types of classification problems</li></ul> | <ul><li>Relies on a predefined distance in feature space</li><li>Slow data processing speed; not suitable for categorical data</li></ul> |
| **IR techniques** | | |
| PageRank (Mohammad et al. 2009) | <ul><li>Low query time and efficiency</li><li>Documents are ranked in decreasing order of their probability</li></ul> | <ul><li>Query independent, all pages come together</li><li>New pages affect the ranking</li></ul> |
| LDA (Huang et al. 2012) | <ul><li>Classifies documents by high-probability topics</li><li>Does not require annotated training data</li></ul> | <ul><li>Poor scalability</li><li>Inefficient to update the model</li></ul> |
| LSA (Huang et al. 2012) | <ul><li>Easier and faster in terms of applying on new data</li><li>Works well on datasets with diverse topics</li></ul> | <ul><li>Representation is dense and not easy to interpret</li><li>Not the best solution for handling non-linear dependencies</li></ul> |
| TF-IDF (Yasunaga et al., 2019) | <ul><li>Easy to compute the similarity of two documents</li><li>Returns highly relevant documents</li></ul> | <ul><li>Since it is based on the bag-of-words, it does not capture the position</li><li>Cannot capture semantics</li></ul> |

**\*Corresponding author's email:** saeed-ul-hassan@itu.edu.pk

The rest of this review is organised as follows. The subsequent section provides the background to citation indexing. Section 3 presents the systematic approach taken to obtaining the publications for this review. Section 4 presents a detailed review of in-text citation analysis and its applications. Section 5 presents the concluding remarks and future research directions.

## 2    Background to Classic Citation Indexing

The bibliometrics field was pioneered by Eugene Garfield. With citation indexes, he introduced new tools for bibliographical research (Garfield, 1955, 1956). An early assessment of the role of citations in science communication was undertaken by Salton (1963). He discussed their role as a pointer to another publication and confirmed that they are a useful supplement to keywords in identifying relevant documents. Salton (1963) showed that integrating the textual and citation information of a publication significantly improves the performance of a retrieval system. Garfield (1965) noted the importance of classification schemes for citations, arguing that authors may have various motivations, as shown in Table 3. According to Voos and Dagaev (1976), the number of times that a particular publication is cited in an another publication is an indication of its subject relevance to that publication. An analysis by Bonzi (1982) of a cited work across many subjects and types of publication attempted to predict the similarities between the cited and the citing publication from the citation's location in the text, and found that citing a document might serve various functions. Prabha (1983) carried out an empirical study on citation behaviour and found that only one-third of all cited sources are considered important by the authors citing them, and that the way in which authors use citations varies by both discipline and specialty (Hurt, 1987).

Many challenges are associated with manual citation context and content analysis as well as citation categorisation. First, in many studies, individuals who are not experts in the area of reported research provided subjective judgements in performing the manual analysis. Thus, unreliable results may be obtained. Second, the categorisation of a citation by using the anchor text around the citation in full-text involves considerable time and human effort (Pride and Knoth, 2017). Third, a citing sentence can contain multiple references and, even though it may include explicit reference to the target work, another part of the same sentence may not refer to that work at all (Jha et al., 2017). Fourth, a citation may appear several times in a publication; therefore, the citation of a specific publication can have more than one function and can be categorised differently (Erikson and Erlandson, 2014).

**Table 3.** Reasons for citing an article (Garfield, 1965: 85)

| No. | Reason |
|-----|--------|
| 1 | To pay tribute |
| 2 | To give credence |
| 3 | To identify techniques and equipment |
| 4 | To provide a literature review |
| 5 | To correct individual work |
| 6 | To correct the work done by others |
| 7 | To disapprove of the work done by others |
| 8 | To verify a claim |
| 9 | To announce an imminent work |
| 10 | To provide a lead to poorly indexed or uncited work |
| 11 | To verify classes of fact and data |
| 12 | To identify the earliest publication that defined the concepts |
| 13 | To identify the earliest publication that explained an eponymous term or idea |
| 14 | To deny the work and approach of others |
| 15 | To spread the property-related claims of others. |

Since manual analysis is a time-consuming and tedious task (Bertin, Atanassova, Gingras, et al., 2016), citation content and context studies have tended to be performed on small datasets (Bornmann et al., 2018). Moravcsik and Murugesan (1975) proposed a manual citation classification scheme for citations in the field of physics and concluded that 40% are merely acknowledgements. This scheme was also used by Chubin and Moitra (1975) who broke it down into fewer classes; their results indicated that the number of negational citations were high for a short period right after publication, but then decayed quickly. The study concluded that 80% of citations in the field of science studies were to confirm a statement or to point to further relevant information. Oppenheim and Renn (1978) used the same scheme to analyse why some old publications continued to be cited, finding that a high number of citations is associated with both their authors' public profile and writing skills. While the systematic work to classify citations on the basis of personal judgement has begun with Frost (1979), Finney (1979) was among the originators of automated systems. Finney (1979) classified medical literature using seven categories and concluded that classification should be based on a citation's location and the cue words (terms) around the citation in the full-text. Small (1982) examined how citations are used in the citing publications and argue that not all citations are of equal importance.

*Corresponding author's email: saeed-ul-hassan@itu.edu.pk

In recent years, access to full-text scholarly publications allows the scientific community to extract various features of a citation, particularly those relating to its function and purpose (Abu-Jbara et al., 2013; Siddharthan and Teufel, 2007), location (Boyack et al., 2018), polarity (Hatzivassiloglou and McKeown, 1997), and linguistic pattern (Ikram and Afzal, 2019). Some studies have used publications in XML format to develop classifiers to identify a citation's function, purpose, and polarity (Jha et al., 2017), thus, demonstrating that the analysis of large-scale datasets and feature extraction (citation context, citation location, sentiment, etc.) have become considerably easier and faster (Hu et al., 2013). These developments have led to new areas of research, such as the use of citation contexts for creating summaries of scientific publications (Cohan and Goharian, 2018; Hoffmann and Pham, 2003) and improving scholarly recommendation and retrieval systems (Doslu and Bingol, 2016; Fang, 2017; Zarrinkalam and Kahani, 2013).

## 3   Data Collection

In order to conduct our literature overview, we adopted a systematic approach. We started with seed articles (publications from our previous research on the topic) and, by reading them, compiled a list of the candidate keywords. On the basis of these keywords, we constructed a complex query to search for the relevant literature in Scopus: title-abs-key("automatic citation classification") or title-abs-key("automated citation classifier") or title-abs-key("automatic indexing" and "citation analysis") or title-abs-key("automatic feature selection" and "citation analysis") or title-abs-key("bibliographic reference classification") or title-abs-key("citation classification") or title-abs-key("citation context") or title-abs-key("citation polarity") or title-abs-key ("citation relation") or title-abs-key("citation-based summaries") or title-abs-key("content citation analysis") or title-abs-key ("content-based citation analysis") or title-abs-key("contextual information" and "citation analysis") or title-abs-key ("scientific citation classification").

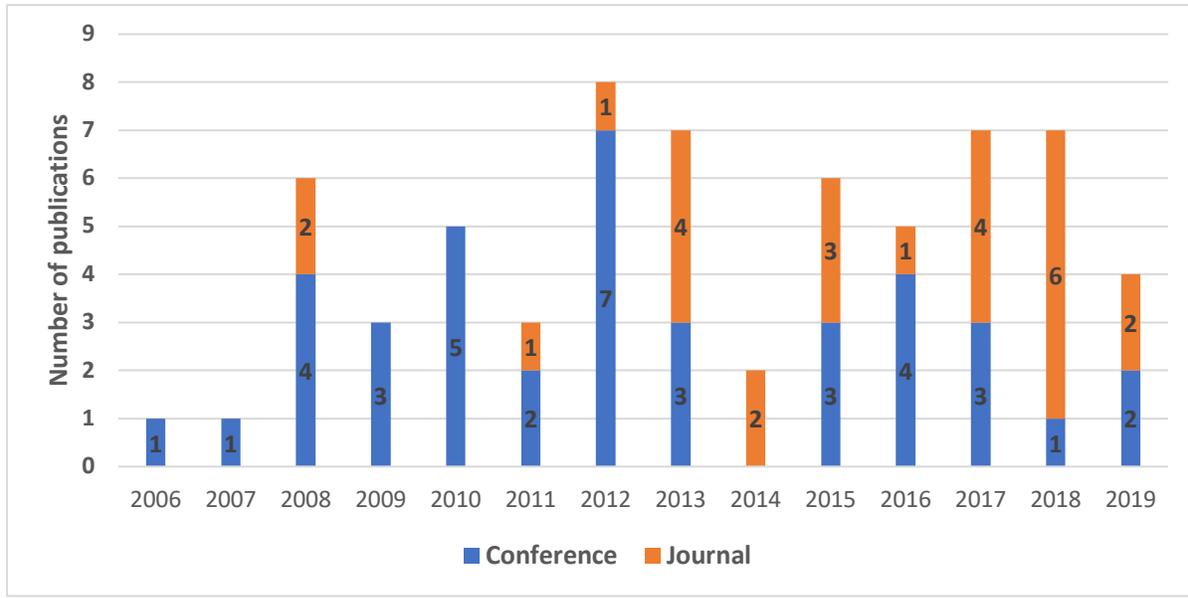

**Figure 1:** Year-wise distribution of selected publications

The above query returned 368 publications that were indexed in Scopus from 2005 to 2019. We combined all refereed documents in these publications with our seed publications. The extended dataset of over 400 publications was then screened to exclude any irrelevant publications, by reading through their titles and abstracts. For instance, the paper 'Scholarly networks on resilience, vulnerability, and adaptation within the human dimensions of global environmental change' were among the most-cited publications (cited 274 times) in our dataset. We discarded it, yet, as it does not focus on the process of citation. Of the remaining publications, as we sought to review only those publications that employ NLP or ML (the focus of this review) and were published after 2005, we removed all irrelevant publications by reading through the abstract of each publication. We finally left with a set of 65 publications, of which 26 were articles published in journals, and the remaining 39 were conference papers. This reflects that conference venues mostly drive the research in the area of in-text citation analysis using NLP and ML methods. We also observed that of the 65 selected publications, journal articles had been slightly increased in recent years (see Figure 1).

Furthermore, to perform keyword- and author-level analysis on our publication set, we needed information pertaining to the authors, citation counts, venues, etc. To retrieve this information, we searched for the Scopus ID of each publication and constructed a Scopus query to yield the required information. Figure 2 shows the overall systematic approach taken by this review.

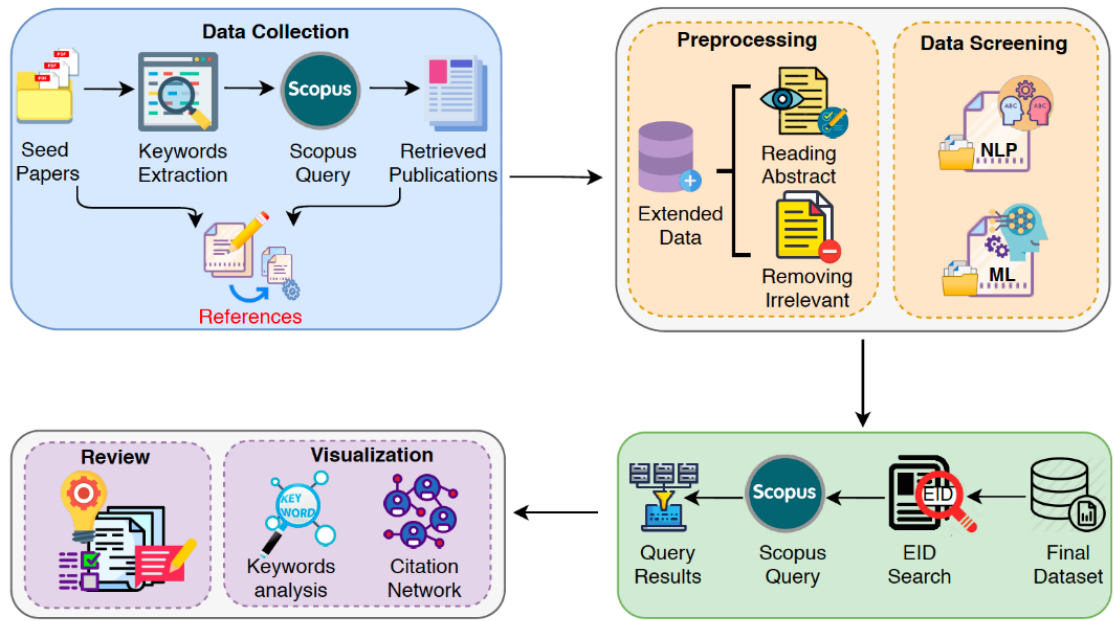

**Figure 2.** A systematic approach for data collection and reporting of the results

To obtain an overview of the active authors in the field of citation context and content analysis over the years, we analysed the citation patterns by using the CitNet Explorer software (www.citnetexplorer.nl; van Eck and Waltman, 2017). In order to locate the major players, we imposed a threshold whereby each publication had to have a minimum of two citations. Figure 3 shows the authors' network of citing publications based on their citation score. Each circle in the network represents a publication's author (since there may be multiple authors of a publication, however, we used only the first author's last name) and is connected to other authors via a direct citation link.

As Figure 3 shows these authors' works form three major clusters that are constructed using internal citation links (based on the publications in our dataset). Eugene Garfield, the founder of the field of bibliometrics, is located at the top of the network. The studies (in light-blue and purple) are the initial studies on citation analysis, such as those by: (1) Garfield (1955), who presents the reasons for citing publications and a classification scheme for citing behaviour; (2) Merton (1957), the founder of modern sociology of science, who basically states that, by citing their work, scholars pay tribute to colleagues; (3) Lipetz (1965), who examines the relationship between citing and cited documents and identifies several citing reasons; and (4) Spiegel-Rosing (1977), who conducts a citation context study using articles from *Science Studies*. The cluster on the right side of the network (in blue) pertains to human-assisted qualitative studies. These include: Small and Greenlee's (1980) explanation of the significance of citation context, paving the way for co-citation analysis and semantic studies of shared knowledge in the scientific domain; Peritz (1983) who stated that the role of citations varies across the disciplines, proposing an eight-category assessment of citation quality; and Conrad and Dabney (2001) who suggested the automatic classification of citation styles (i.e., IEEE, APA, etc.), which professionals had previously identified manually.

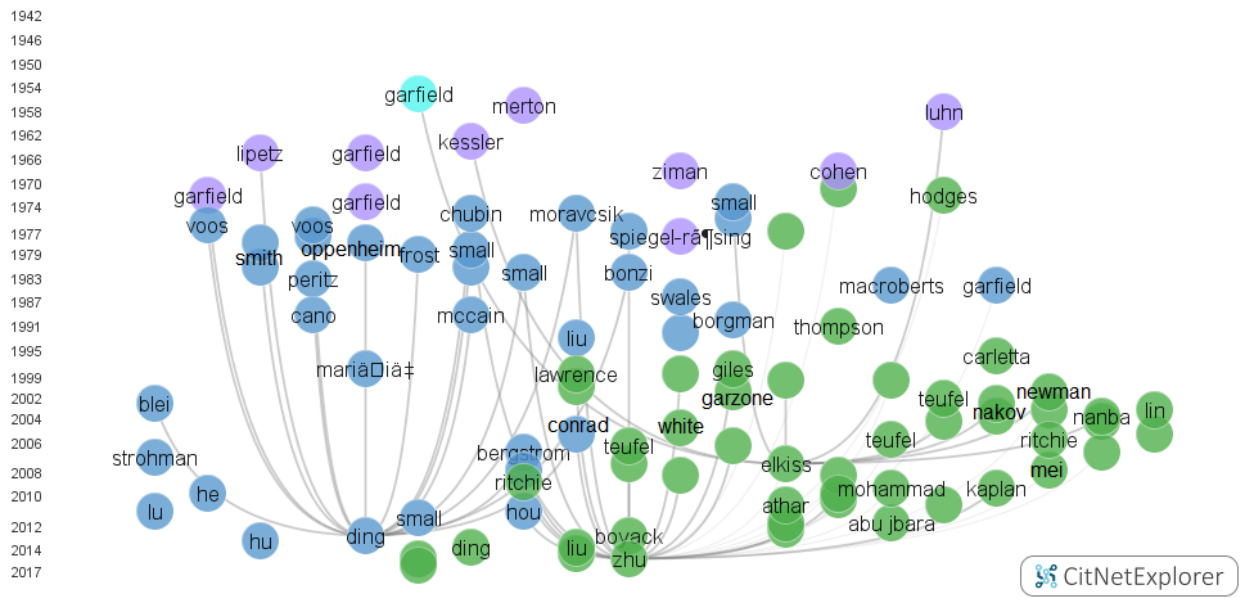

**Figure 3.** A network of authors associated with citation content and context analysis

The circles on the right (in green) represent authors who focused on the automated analysis of a citation's context. In order to understand the relationship between cited and citing publications, Nanba and Okumura (1999) propose using cue phrases (phrases around a citation, including adverbs, pronouns, negative expressions, etc.) for the extraction and classification of citation contexts. Garzone and Mercer (2000) propose an automated classification scheme that consists of 35 classes, based on a citation's cue words and section in the article in which it appears. Nakov, Schwartz, and Hearst (2004) found that the citation context aligns very well with rich information and can be used for document summarisation, training, and testing data for semantic analysis and information retrieval. In this review, we are especially interested in the green nodes and links.

To examine the keywords associated with the publications in our set, a network based on author-defined keywords has been constructed using VOSviewer (www.vosviewer.com; van Eck and Waltman, 2017) to visualise frequent keywords. We have selected 79 terms, applying a minimum of two occurrences as a threshold so that we focus on the major research streams. The size of the nodes is determined by counting the total occurrences, while the co-occurrence of terms determine the width of the links: the higher the co-occurrence, the thicker the line. Figure 4 shows a detailed view of the research landscape of citation content and context analysis, presenting the terminologies in this field. The clusters of associated keywords appear in the network, indicating their relatedness in terms of relevant articles.

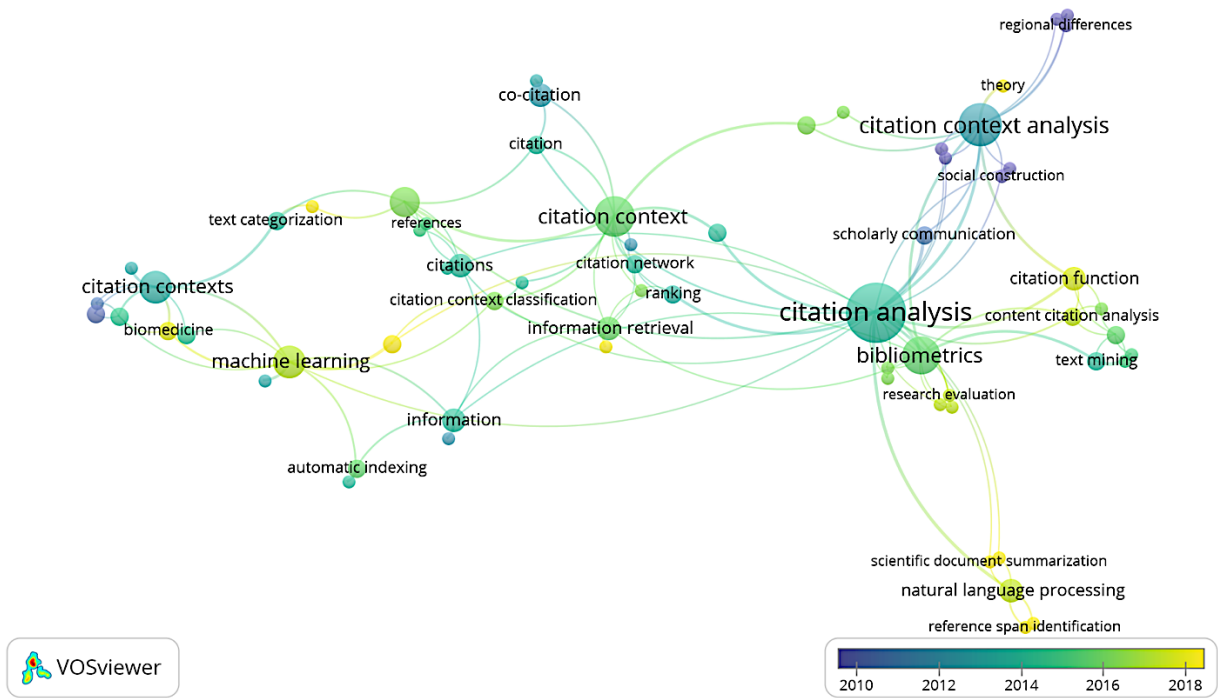

**Figure 4.** Visualisation of author keywords over time using VOSviewer

To visualise the evolution of this research area, a time window indicates the keywords used over the years (since 2009). Upon examining the 'citation analysis' cluster, it can be seen that it deals with (1) content-based citation analysis, (2) citation context analysis, (3) identifying citation functions, (4) implying schemes for citation context classification and the use of n-grams as a feature for citation classification, (5) building citation-based networks, (6) information retrieval, and (7) applying NLP techniques to design citation recommendation and ranking systems. Relatively new topics in this research area are the measurement of knowledge flows between academic institutions and authors, the automatic classification of citations using NLP and ML techniques, the design of context-aware citation recommendation systems, and the discrimination between influential and non-influential citations.

In this literature review, we are especially interested in new advancements that leverage NLP and ML approaches in citation context, citation content and citation sentiment analysis, citation-based summarisation, and citation-based retrieval systems. These topics were not in the focus of previous reviews on citation content and context analysis.

## 4   Review of In-text Citation Analysis

Due to the various citation formats and styles, the identification of a citation anchor in a publication is a difficult and challenging task (Cronin, 1982; Shahid et al., 2015). In the past, many methods have been proposed to detect in-text citations, such as CiteSeer (Giles et al., 1998) and Cora (McCallum et al., 2000). Councill et al. (2008) pioneered an open-source package (ParsCit) to locate reference strings. ParsCit employs ML methods coupled with a heuristic framework to find and delimit reference strings, locate the context of a citation, and parse the document structure accurately. Lopez (2009) launched the online tool GROBID (GeneRation Of BIbliographic Data) to identify and parse the citation anchors in citing documents. Tkaczyk and Bolikowski (2015) invented the tool CERMINE (Content ExtRactor and MINEr) to extract the context of citations.

The automatic detection of in-text citation anchors has led to many new research areas. Based on the various research problems, methods, and techniques found in the studies of our publication set, we have divided this review into five distinct parts: citation content and context analysis, citation classification, citation-based sentiment analysis, citation-based summarisation, and citation recommendation systems. We discuss the corresponding studies in the following sections. A summary of the reviewed articles can be found in Table 4.

**Table 4.** Summary of the reviewed literature on citation in-text analysis

| Article | Data Repository | Sample Size | Main Results |
|---|---|---|---|
| **Citation context and content** | | | |
| Ritchie, Robertson, and Teufel (2008) | Association of Computational Linguistics (ACL) Anthology | 9,800 papers | Longer citation contexts resulted in greater retrieval effectiveness, 3sent was more effective than 1sent. |
| Angrosh, Cranefield, and Stanger (2010) | Lecture Notes in Computer Science (LNCS) | 50 papers | Citation features along with the sentence features play an important role in the identification of citation context and yield an accuracy of 96.51%. |
| Aljaber et al. (2011) | TREC Genomic | 162,259 papers | Citation context is a rich source of topically related terms and many of them are semantically related to terms that are present in the original document. |
| Angrosh, Cranefield, and Stanger (2013) | Lecture Notes in Computer Science (LNCS) | 20 papers | CRF with additional zero-order features identified context better than linear CRF and scored an accuracy of 91%. |
| Zhang, Ding, and Milojević (2013) | Social Sciences | - | Citations should not be treated equally as they have different reasons and functions. |
| Hu, Chen, and Liu (2013) | *Journal of Informetrics* | 350 papers (11,327 citations) | In full-text articles, citations are distributed unevenly, more than 50% of the citations belong to the introduction section. |
| Ding et al. (2013) | *Journal of the American Society for Information Science and Technology (JASIST)* | 866 papers | Highly re-cited references (the same publication is cited multiple times in the citing paper) appear mostly in the introduction and literature review sections. |
| Abu-Jbara, Ezra, and Radev (2013) | ACL Anthology | 30 citing papers (3,500 citation contexts) | In the interpretation of citation contexts, lexical features (determiners and conjunction adverbs) are more significant than structural features (position and reference). |
| Bertin and Atanassova (2014) | Public Library of Science (PloS) Journals | 9,446 papers (459,834 citation sentences) | The frequency of verbs in citation contexts depends on the paper sections in which they appear and 50% of the verbs were present in the introduction section. |
| Fujiwara and Yamamoto (2015) | PMC-OAS | 545,147 papers | Papers cited in less than five citation contexts account for around 76% of all the cited papers in the database. |
| Hu, Chen, and Liu (2015) | *Journal of Informetrics* | 350 papers (11,327 citations) | In research articles, succeeding citations are more intentional and purposeful than first-time citations. |
| Bertin, Atanassova, Sugimoto, and Lariviere (2016b) | PLoS Journals | 75,000 citing papers (3 million citation sentences) | The word 'show' is the most frequently occurring verb in citation contexts among all paper sections. |
| Bertin and Atanassova (2017) | PLoS Journals | 80,000 papers (3,528,514 citing sentences) | 41% of the citation sentences contain MIR, most of them appear in the introduction section. |

| Small, Tseng, and Patek (2017) | PubMed Central Open Access Subset (PMC-OAS) | 1.1 million papers | Only 46% of the articles that had 'discovery' words in their citing sentences (citances) were actually scientific discoveries. |
|---|---|---|---|
| Hu et al. (2017) | *Journal of Informetrics* | 350 papers (16,917 citations) | 25% of the references were mentioned multiple times and located in close proximity. |
| Boyack et al. (2018) | Elsevier and PubMed Central | 5 million papers | The references that are mentioned just once are typically more highly cited than references that are mentioned multiple times. |
| **Citation classification** | | | |
| Teufel, Siddharthan, and Tidhar (2006) | ACL Anthology | 360 papers | The proposed annotation scheme achieved 77% accuracy in determining citation functions. |
| Siddharthan and Teufel, (2007) | Computation and language (CmpLg) | 320 papers | Scientific attribution features play a vital role in determining the citation functions and achieved an accuracy of 74.7%. |
| Sugiyama, Kumar, Kan, and Tripathi (2010) | ACL Anthology Reference Corpus (ARC) | 10,921 papers (112,533 citing sentences) | Proper nouns and the previous and next sentence features are more effective than uni-grams and bi-grams for citation classification. |
| Agarwal, Choubey, and Yu (2010) | PubMed Central | 1,710 citing sentences | N-grams play an important role in analysing the function of citations with an accuracy of 92.2%. |
| Dong and Schäfer (2011) | ACL Anthology | 1,768 citing sentences | Syntactic patterns that extracted from Part of Speech (POS) tags are most effective for achieving good classification results. |
| Jochim and Schütze (2012) | ACL Anthology Reference Corpus (ARC) | 84 papers | Lexical features play a vital role in classifying citations and alone achieved a micro F score of 0.61. |
| Wang, Villavicencio, and Watanabe (2012) | *IEEE Transactions* | 40 citing papers (345 citation contexts) | The highest frequency of citation contexts belongs to the extend class (50%). The next classes were to criticise (30.14%), compare (13.88%), and improve (3.83%). Cue phrases identify the relationship between cited and citing articles accurately. |
| Abu-Jbara, Ezra, and Radev (2013) | ACL Anthology | 30 citing papers (3,500 citation contexts) | The occurrence of the used class is highest (14.7%) followed by criticism, substantiation, and basis classes. |
| Zhu, Turney, Lemire, and Vellino (2015) | ACL Anthology | 100 citing papers (3,143 papers cited) | The total number of times a paper was cited in the citing paper was the most predictive feature for impact. |
| Valenzuela, Ha, and Etzioni (2015) | ACL Anthology | 465 cited–citing paper pairs | Most citations in papers are non-influential and only a small proportion (14.6%) is influential. |
| Hassan, Akram, and Haddaway (2017) | ACL Anthology | 465 cited–citing paper pairs | The most significant feature to identify the important cited papers for the citing papers is the 'similarity between the cited paper's abstract and the text around the citation in the body of the citing paper'. |
| Jha et al. (2017) | ACL Anthology | 30 citing papers (3,500 citations) | The SVM classifier outperformed the naïve bayes and logistic regression classifiers in determining citation functions with an accuracy of 70.5%. |
| Pride and Knoth (2017) | ACL Anthology | 415 cited–citing paper pairs | The combination of three features (total number of direct citations, author overlap, and abstract similarity) led to good classification results. |
| Small (2018) | PubMed Central | 1,000 most-cited papers in PubMed (646,347 citances) | The frequency of the hedging word 'use' is higher in methods papers, whereas the frequency of other hedging words (such as 'may,' 'show,' and 'suggest') is higher in non-methods papers. |
| Hassan, Imran, Iqbal, Aljohani, and Nawaz (2018) | ACL Anthology | 465 cited-citing paper pairs | Deep learning-based LSTM models perform exceptionally well in identifying the important citations for a given article. |
| Hassan, Iqbal, Imran, Aljohani, and Nawaz (2018) | ACL Anthology | 465 cited-citing paper pairs | Using unsupervised techniques, a large cluster of non-important citations is obtained. |
| Tuarob et al. (2019) | CiteseerX | 8,796 citation contexts | For binary algorithmic citation classification, heterogeneous ensemble methods achieved the best average F1 score. |
| **Citation sentiment analysis** | | | |
| Teufel, Siddharthan, and Tidhar (2006) | ACL Anthology | 360 papers | A three-way classification scheme achieved an 8 percentage points better results than a four-way classification scheme. |
| Piao, Ananiadou, Tsuruoka, Sasaki, and McNaught (2007) | PLoS Journals | _ | Polarity relations between cited and citing documents are useful for information retrieval and text-mining. |
| Athar (2011) | ACL Anthology | 310 papers (8,736 citations) | Dependencies' features and 3-grams achieved the best results (0.89 macro F1) for the task of automatic identification of sentiments. |
| Athar and Teufel (2012a) | ACL Anthology | 310 papers (8,736 citations) | Sentiment analyses achieve good results with a context window of four sentences. |
| Athar and Teufel (2012b) | ACL Anthology | 310 papers (8,736 citations) | Four-sentence context is better than single-sentence in sentiment analysis; the reduction leads to a loss of much of the sentiment. |
| Li, He, Meyers, and Grishman (2013) | PubMed Central | - | A class imbalance problem in the data corpus leads to a poor performance of polarity classifiers. |
| Abu-Jbara, Ezra, and Radev (2013) | ACL Anthology | 30 citing papers (3,500 citation contexts) | The use of contexts improved the polarity systems' accuracy significantly by 10 percentage points. Features associated with subjectivity, such as negation, subjectivity cues, and speculation, are important for the classification of polarity. |
| Sula and Miller (2014) | *Art Bulletin*, *Language*, *Journal of Philosophy*, PMLA | 37 papers, 30 papers, 18 papers, 77 papers respectively | Most of the contexts belong to the neutral class, as they don't clearly demonstrate the polarity. Negative and positive training sets should be larger to increase the accuracy of polarity classifiers. |
| Jha et al. (2017) | ACL Anthology | 30 citing papers (3,500 citations) | Eliminating the neutral class increased the accuracy by up to 6 percentage points. |
| Ikram, Afzal, and Butt (2018) | ACL Anthology | 310 paper (8,736 citation) | Higher values of n-grams yielded better scores in automatically identifying citation sentiments than lower values. |
| Taşkın and Al (2018) | Turkish LIS Publication | 423 documents (101,019 citing sentences) | Increasing the number in the negative class training set improves the accuracy. |
| Ikram and Afzal (2019) | ACL Anthology, | 310 paper (8,736 citation), 285 papers | A high value of n-grams (about n = 5) achieves the best results for sentiment classification. |

| | Clinical Trial Papers | | |
|---|---|---|---|
| **Citation-based summarisation** | | | |
| Elkiss et al. (2008) | PubMed Central | 2,497 papers | Citing sentences contain additional information for automatic summarisation which can be used as a substitute in the absence of abstracts. |
| Mei and Zhai (2008) | Medical Literature Analysis and Retrieval System Online (MEDLINE) | 14 papers | Language-based models are effective and outperform baseline methods, i.e., MEAD, in summarisation. |
| Qazvinian and Radev, (2008) | ACL Anthology | 11,000 papers | The C-lexrank model attained 6 and 4 percentage points higher nugget based-pyramid scores than the C-RR and LexRank models. |
| Mohammad et al. (2009) | ACL Anthology | 11,000 papers | Both abstract and citation contexts contain valuable information, effective for the summarisation task. |
| Kaplan, Iida, and Tokunaga (2009) | Computational Linguistics | 38 papers citing 4 papers | The identification of citation contexts using the co-reference analysis approach achieved F1-scores of about 85% for research paper summarisation. |
| Qazvinian and Radev (2010) | Computational Linguistics | 10 papers | Summaries based on contextual information and citing sentences have higher quality (Pyramid score 0.63) than those generated by using citing sentences alone (Pyramid score 0.41). |
| Abu-Jbara and Radev (2011) | ACL Anthology | 55 papers (4,335 citation sentences) | Sentence filtering has a significant impact on the summarisation results and a proposed technique achieved a 0.53 ROUGE score. |
| Tandon and Jain (2012) | Microsoft Academic (MA) | 30 papers (500 citation contexts) | Adjective, verb, and bi-gram models beat the accuracy of the LM model. The LM accuracy could be improved by cleaning and increasing the data. |
| Barrera and Verma (2012) | DUC 2002 | 533 papers | The position model prioritising sentences closer to headings achieved better results than TextRank and MEAD, because sentences close to the headings are likely to be more content representative for summarisation. |
| Qazvinian et al. (2013) | ACL Anthology | 30 papers | The generation of technical summaries benefits considerably from the use of citing sentences. |
| Conroy and Davis (2015) | Biomedical Summarisation data | 20 documents (referenced papers) with 10 documents (citation papers) | The term frequency vector space model is a good basic model, but the nonnegative matrix factorisation exceeds the ROUGE scores of computed summaries. |
| Klampfl, Rexha, and Kern (2016) | Computational Linguistics | 40 annotated sets of citing and reference papers | The TextRank algorithm performed best in extracting the 'most relevant' key terms and sentences for summarisation. |
| Nomoto (2016) | Computational Linguistics | 4,608 training instances | The performance of summarisation models may lack due to the inability to clearly define false and true targets. |
| Lu, Mao, Li, and Xu (2016) | Computational Linguistics | 40 annotated sets of citing and reference papers | For facet identification, SVM achieved the best average performance (F1=0.65). Naïve Bayes tackled the class imbalance problem efficiently and showed balanced performance on each facet. |
| Cao, Li, and Wu (2016) | Computational Linguistics | 40 annotated sets of citing and reference papers | The distribution of facet among the data is highly unbalanced, 60% belong to the method facet while only 9% belong to hypothesis. |
| Verma and Lee (2017) | DUC 2002 | 533 papers | System generated summaries are needed because they contain 9 percentage points more related keywords from the original document than human abstractive summaries. |
| Ma, Xu, and Zhang (2018) | Computational Linguistics | 40 citing and cited papers set | Similarity-based features are more suitable for summarisation than position-based features. |
| Yasunaga et al. (2019) | Computational Linguistics | 1,000 papers | Incorporating both the community's views and authors' original insights about the article for summarisation lead to better results than traditional citation-based and abstract-based summaries. |
| **Citation-based recommendation system** | | | |
| Nallapati, Ahmed, Xing, and Cohen (2008) | CiteSeer | 3,312 papers | Pairwise Link-LDA is more expressive than Link-PLSA-LDA in terms of modelling an arbitrary link structure. |
| Tang and Zhang (2009) | NIPS and CiteSeer | 1,605 papers and 3,335 papers | Two-layer restricted Boltzmann machine model performed better than LM-based techniques for citation recommendation but the data sparsity could be an issue in achieving the highest accuracy. |
| He, Pei, Kifer, Mitra, and Giles (2010) | CiteSeerX | 1,810,917 citation contexts | Citation counts slightly increased the effectiveness of citation recommendation systems compared to uni-grams, bi-grams, and dependency models. |
| Wang and Blei (2011) | CiteULike | 37 papers | Collaborative topic regression models have good predictions on completely unrated articles in comparison to traditional topic modelling approaches like LDA. |
| Huang et al. (2015) | CiteSeer | 1,017,457 papers (10,760,318 citation contexts) | The probabilistic neural network model improves the overall recommendation with a 9 percentage points gain on recall compared to the LDA. |
| Gupta and Varma (2017) | Arnetminer | Training (74,097 papers), testing (618 papers) | The combination of distributed representations of scientific article's content and graph improves a recommendation system accuracy by 8 percentage points on recall as compared to TFIDF. |
| Yang et al. (2018) | ACL Anthology, DBLP | Training (11,197 papers), testing (1,358 papers) | Parameters of weights may influence the performance of context-aware citation recommendations. |

## 4.1 Citation Context and Content

The extraction of citation context is a vital task for studying the various facets of the relationship between citing and cited publications. Good access to the data is essential. Fujiwara and Yamamoto (2015) constructed a Colil (Comments on Literature in Literature) database containing extracted citations and co-citations from 545,147 full-text articles from PubMed Central Open

Access Subset (PMC-OAS). They used a newly compiled vocabulary and the Resource Description Framework (RDF) – a technology for publishing, describing, and linking life sciences data on the web. They developed a web-based search service for a cited article. This service returns details of the citation context along with the article's co-citations. The results from Colil were compared to Microsoft Academic Search (MAS: another system for extracting citation contexts at that time). In response to a keyword search, the Colil system had a higher number of indexed articles than MAS in 2015. More recently, however, the MAS (now Microsoft Academic) has become a powerful system with about 233 Million publications, which questions the use and need of other systems like Colil.

### 4.1.1 Citation Context Window Size

Studies have presented varying views about the extent of the context relevant to analysing the connection between the cited and citing publication. Ritchie, Robertson, and Teufel (2008) compared various sizes that facilitate the application of ML techniques. They used the following nine categories to define citation context: none (contains no citation context), 1sent (contains only the citing sentence); 3sent (contains the citing sentence plus one sentence before and one after); 1sentupto (contains one sentence context, truncated at the next citation); 3sentupto (contains three sentences context, truncated at the next citation); win50 (contains 50 words on the left and right of a citation); win75 (contains 75 words on the left and right of a citation); win100 (contains 100 words on the left and right of a citation); and full (contains the full citing paper). Ritchie, Robertson, and Teufel (2008) assumed that words that describe the cited publication are located close to the citation in the full-text than words that are far away from the citation.

The results of the study showed that the 3sented context performed better than 1sent, that 1sentupto and 3sentupto performed worse than 1sent and 3sent, respectively. win50 performed worse than either win75 or win100. A comparison of the effectiveness of window-based and sentence-based contexts proved the greater usefulness of sentence-based contexts; expanding the context size did not ensure any better identification of the contexts. In a succeeding study, Athar and Teufel (2012a) suggested the use of four sentences (the citing sentence, one sentence before the citing sentence, and two sentences following the citing sentence) as an appropriate citation context window. According to the authors, the longer citation contexts (four sentences) are more effective than shorter citation contexts (only the citing sentences), as the longer citation contexts contain more descriptive terms from the citing article. The four-sentence context window is now considered as a quasi norm for citation context analysis studies (Teufel, Siddharthan, and Tidhar, 2006).

### 4.1.2 Feature Extraction for Context Identification

In automatic context identification, feature extraction is a non-trivial and vital component in the training of ML models. Angrosh, Cranefield, and Stanger (2010) described the need for contextual aspects by introducing the eleven sentences and two citations features. In an experiment, they applied the conditional random fields (CRF) algorithm to a data corpus of 50 articles from lecture notes in computer science. The results showed that the use of sentence features in labelling sentences achieved an encouraging accuracy of 93.22%. Citation features that denote the presence of citations play an essential role, and their use achieved an enhanced accuracy of 96.51% and 74.3% F1 measure scores. The F1 measure is a weighted harmonic mean of the precision (percentage of total results which are relevant) and recall (percentage of relevant results correctly classified by the algorithm).

**Table 5.** Features used for identifying citation context (Abu-Jbara, Ezra, and Radev, 2013 : 599)

| Feature | Description |
| --- | --- |
| Demonstrative determiners | The current sentence contains a demonstrative determiner (this, that, these, etc.) |
| Conjunctive adverbs | The current sentence starts with a conjunctive adverb (however, accordingly, furthermore, etc.) |
| Position | Position of the current sentence with respect to citation |
| Contains closest noun phrase | The current sentence contains the closest noun phrase (method, corpus, or a tool) |
| 2-3 grams | The first bi-gram and tri-gram in the sentence contain references other than the target |
| Contains mention of the target reference | The sentence contains a mention (explicit or anaphoric) of the target reference |
| Multiple references | The citing sentence contains multiple references |

Angrosh, Cranefield, and Stanger (2013) used a similar dataset as Angrosh, Cranefield, and Stanger (2010) and compared the first-order linear chain CRF approach with the CRF approach with additional zero-order features. The new study found that CRF with additional zero-order features performed better because of its back-off prediction capability. Moreover, they recognised an optimal set of features for the identification of citation context and found that among two citation features, the 'prevSentHasCitation' feature was insignificant. Removing this feature left the overall accuracy unchanged. A study by Abu-Jbara, Ezra, and Radev (2013) treated automatic context identification as a supervised sequence-labelling problem and computed seven novel features by applying the CRF technique. They experimented with a data corpus of 30 full-text articles from the Association of Computational Linguistics (ACL) and proved that the CRF approach outperformed all baseline models in identifying a citation's context with a precision of 98.5%, 82% recall, and 89.5% F1 measure score. With respect to the best performing features, their findings contradicted those of Angrosh, Cranefield, and Stanger (2010). The authors showed that lexical features (determiners and conjunction adverbs) are more significant than structural features (position and reference). The F1 measure score achieved by Abu-Jbara, Ezra, and Radev (2013) is 15 percentage points higher than that of Angrosh, Cranefield, and Stanger (2010). The results point out, therefore, that their feature set (see Table 5) should be preferred in future studies for the task of identifying citation contexts.

### 4.1.3 In-text Citation Distribution

Several studies have focused on the distribution of in-text citations by considering the IMRaD structure (i.e., introduction, methods, results, and discussion) in which a citation may appear. Hu, Chen, and Liu (2013) focused on the distribution of citations across the sections, by analysing a dataset of 350 XML-formatted publications in the *Journal of Informetrics*. The distribution and density of citations were 1,285 (41.8%) in the introduction, 776 (25.2%) in the methods, 796 (25.9%) in the results, and 271 (7%) in the discussion section. In other words, the citation density in the introduction is significantly higher than that in other sections. The results of the study suggested that if a publication has more citations in the methods section, the focus of the article is on methodology. Whereas if an article has an even distribution of citations across the sections, there is a high probability that it is a review. Another study focusing on section-specific citation distributions has been presented by Ding et al. (2013). They used a larger dataset than Hu, Chen, and Liu (2013) consisting of 866 articles from JASIST. The authors included only highly cited references and used structural and semantic features in the analyses. They reported that the use of citations in the documents is unequal: some cited references appear multiple times and some only once in the text. The study reported that 78% of citations were cited less than three times, and the most highly cited publications are found in the introduction and related work sections.

In a succeeding study, Hu, Chen, and Liu (2015) also investigated the phenomenon of recurring citations. Among the total of 11,327 citations from 350 articles, 74.3% (8,417 citations) were cited only once in the citing publication, 25.7% (2,910 citations) were cited twice or more. The study examined the frequency of citations in two perspectives: citation context and citation location. In citation location analysis, they determined the location distribution of recurring citations by using the IMRaD structure and found that the most cited reference was cited in a similar section. This result revealed that a reference seems to be cited within a single topic or context. For the citation context analysis, Hu, Chen, and Liu (2015) extracted the context of first time and succeeding cited publications and found that first-time citations are perfunctory. Since succeeding citations were frequently more purposeful, authors just mentioned a citation in the beginning and then explained it meticulously when they cited it again. A challenge for the study was the various used styles of citations in the papers (IEEE, APA, and Harvard); it is difficult for a model to differentiate between them. Many results by Hu, Chen, and Liu (2015) could be confirmed by a recent study of Hu et al. (2017), who used the dataset of 350 papers from the *Journal of Informatics.* The authors stated that 25% of the cited references were mentioned multiple times in a citing publication. Multiple mentioned references were cited in similar sections or in close proximity.

Boyack et al. (2018) investigated 5 million publications from Elsevier and PubMed Central to mine the characteristics of in-text citations. They showed that the distribution of citations among the sections of scientific publication is even; except in the methods section that showed more recurring citations. This result might contradict the findings of Ding et al. (2013) which indicated that most of the highly cited works can be found in the introduction and literature review sections. The study also found that references that are cited only once in the publication are more frequently highly cited than references that are mentioned multiple times. The reason is that citations cited only once are usually older and thus have higher citation counts. Bertin and Atanassova (2018) investigated full-text articles for multiple in-text references (MIR) and their locations. Various NLP-based techniques were used to extract MIR from a large dataset of 80,000 publications from the Public Library of Science (PLoS). The findings indicated that: a) MIR frequently appear in all sections (about 41% of the sentences with citations); b) MIR appear quite often in the introduction, discussion, and results sections (about 20% of the sentences), and less so in the methods section (only 15% of the sentences); c) MIR are mostly found near verbs within a sentence.

The various studies that have investigated in-text citation distributions hitherto have used very different datasets. They agreed on the point, however, that a considerable part of cited publications are cited multiple times in citing publications and these multiple citations are in (close) proximity (the same or similar sections).

### 4.1.4   Citations' Role According to Position

A handful of studies investigated the roles of citations according to their position in scientific articles. These studies leveraged the terms' or verbs' frequencies appearing in the citation contexts in the IMRaD structure. Aljaber et al. (2011) empirically analysed biomedical terms (animals, cell lines, mutation, etc.) that they found within the citation contexts by using 162,259 biomedical publications from TREC Genomic. They observed that a citation context is a rich source of topically related terms. Many of the terms were semantically related to terms that are present in the citing publication. Aljaber et al. (2011) analysed two different aspects of citation terms: a) the section in which they are present b) and the distance of the term to the citation marker. The authors concluded that the section is related to the quality of citation terms, and most of the citation terms are located in the introduction and discussion sections.

Bertin and Atanassova (2014) worked on the identification and density of the verbs in citation contexts by using a dataset of 9,446 articles from five PLoS journals. After locating the verbs in the citation contexts, they ranked them by frequency in each section and noted that 50% were alone in the introduction section. The word 'show' was the most common verb in both the introduction and the discussion sections, but the second-most frequent verb in the results section. Similar to the finding of Aljaber et al. (2011), the study concluded that citations play a unique role depending on their position in the structure of an article. In a succeeding study, Bertin, Atanassova, Sugimoto, and Lariviere (2016b) extended the dataset (75,000 citing papers) for the identification of linguistic patterns in citation contexts and explored whether these varied by the location of citations (using an NLP approach). Similar to Bertin and Atanassova (2014), the authors observed that the introduction and discussion sections contain most of the verbs; the word 'show' was the most frequently occurring verb among all sections.

In the biomedical domain, Small, Tseng, and Patek (2017) recently explored the citing sentence, called 'citance,' to identify the word 'discoveries' for classifying the publications that present scientific discoveries. The authors used 1.1 million full-text publications from the PubMed Central Open Access Subset (PMC-OAS) database. They trained an ML classifier on the citance vocabulary and found that only 46% of the publications that had the term 'discovery' in their citance were scientific discoveries. The authors concluded that the term 'discovery' is not a reliable feature to find (citing) publications that present discoveries. They also presented the top 10 words associated with discoveries; the top three are 'discovered,' 'first,' and 'important.' Concerning the techniques used for classification, they concluded that the ridge regression classifier performed best, with an accuracy of 94%.

**Table 6.** Annotation scheme for citation function (Teufel, Siddharthan, and Tidhar, 2006: 105)

| Category | Description |
|----------|-------------|
| Weak | Weakness of the cited approach |
| CoCoGM | Contrast/Comparison in goals or methods (neutral) |
| CoCo- | Author's work is stated to be superior to cited work |
| CoCoR0 | Contrast/Comparison in results (neutral) |
| CoCoXY | Contrast between two cited methods |
| PBas | Author uses cited work as the basis or starting point |
| PUse | Author uses tools/algorithms/data/definitions |
| PModi | Author adapts or modifies tools/algorithms/data |
| PMot | This citation is positive about the approach used or problem addressed (used to motivate work in the current paper) |
| PSim | Author's work and cited work are similar |
| PSup | Author's work and cited work are compatible/provide support for each other |
| Neut | Neutral description of cited work, or not enough textual evidence for above categories, or unlisted citation function. |

## 4.2   Citation Classification

Research on the automatic classification of citations has been undertaken by many authors using various classifiers, data corpora, in-text features, and numbers of classes. Among the pioneers, Teufel, Siddharthan, and Tidhar (2006) used a supervised IBK classifier (which is similar to KNN and uses a distance measure to locate k nearest instances in the training data for each test instance) with shallow and linguistically inspired features including cue phrases. The authors used a corpus of 116 documents from the ACL Anthology to present an automated annotation scheme of 12 classes. Table 6 lists these classes. The study shows that the class 'PMot' appears nearer to the beginning of the publications; the comparative result classes (CoCoR-, CoCoR0) mostly appear near the end of publications. The system identifies the class of each citation with an accuracy of 77%. After the publication of this pioneering study, several subsequent studies have revealed various essential features for the task of citation classification.

### 4.2.1 Feature Extraction for Citation Classification

Abu-Jbara, Ezra, and Radev (2013) used a citation taxonomy of six classes by merging the 12 classes proposed earlier by Teufel, Siddharthan, and Tidhar (2006). They experimented with multiple classifiers and employed a battery of features, and found that the SVM classifier outperformed the others, achieving an accuracy of 70.5% in predicting correct citation categories. Although the scheme could not match the accuracy of the classifier by Teufel, Siddharthan, and Tidhar (2006); the study by Abu-Jbara, Ezra, and Radev (2013) revealed the importance of structural (number of references in citation contexts) and lexical (closest verb, adjective, adverb, subjective cue, etc.) features for the classification of a citation's purpose. In a similar way to the approach adopted by Abu-Jbara, Ezra, and Radev (2013), Jha et al. (2017) used Teufel, Siddharthan, and Tidhar's (2006) annotations of a dataset including 3,500 citations from the ACL Anthology. The feature set from the study of Abu-Jbara, Ezra, and Radev (2013) was used to train three supervised classification models (SVM, NB, and LR). The authors reported a similar accuracy (70.5%) as Abu-Jbara, Ezra, and Radev (2013) by using the SVM classifier. Jha et al. (2017) also evaluated each feature's importance. The results agreed to the findings of Abu-Jbara, Ezra, and Radev (2013): both lexical and structural features are vital for citation function classification.

All these studies agreed that among the citation functions, 'used' has the highest occurring frequency and lexical features are essential for the classification of citations. By comparing the accuracy of the various models used in the studies, we recommend using the scheme by Teufel, Siddharthan, and Tidhar (2006) as a benchmark in future studies. The studies have revealed that the performance of the various classification models is not very high; however, increasing the annotated instances in the training of ML models could improve the performance.

Several studies have investigated the most important attributes in determining the functions of citations. Siddharthan and Teufel (2007) proposed four attributive features and evaluated their importance in categorising citations. They designed a seven-category classification scheme and ran five classifiers (HNB, NB, KNN, DT, and STACKING: a combination of NB and DT), with or without attribution features, on a data corpus used by Teufel and Moens (2002). The results depicted that the classifier with lexical, linguistic, and position-based features achieved a macro-F1 value (i.e., the arithmetic mean of per-class F1 scores on different datasets) of 51%. The accuracy was improved by 2 percentage points including scientific attribution features and lexical, linguistic, and position-based features. Siddharthan and Teufel, (2007) also compared their result with the baseline (Teufel and Moens, 2002) and noted that their system achieved a 3 percentage points increased value of macro-F1.

In another study which employed the ACL Anthology dataset, Dong and Schäfer (2011) proposed three novel features: textual (cue words), physical (citation location and density), and syntactic features (part-of-speech 'POS' sequences) to classify citations into four classes: background, fundamental idea, technical basis, and comparison. The classification performance was compared with and without syntactic features using different machine learning classification algorithms. The NB and BayesNet classification models demonstrated the robustness of the proposed features with a micro-F1 value (i.e., a harmonic mean of micro-averaged precision and micro-averaged recall on different datasets) of 64%. Even though the performance of the model was not very well, the study revealed the syntactic patterns in citing sentences: for example, a citing sentence that describes the background of current work is usually in the active voice, while the sentence that introduces the tools or methods used is in the passive voice.

In the most recent study in this area of feature extraction for citation classification, Jochim and Schütze (2012) introduced eight new attributes: lexical features, word-level linguistic features, linguistic structure features, location features, frequency features, sentiment features, self-reference features, and named-entity-recognition features. They built a model to investigate the robustness of the features, using an annotated dataset from the ACL Anthology Reference Corpus (ARC). The authors show that the lexical features alone achieved an encouraging F1 score of up to 61%, whereas, using all features, the model achieved an F1 score of 65%.

Assessing the usefulness of the studies by Siddharthan and Teufel, (2007), Dong and Schäfer (2011), and Jochim and Schütze (2012) for feature extractions is a challenging task, as each of them has used different datasets and annotation schemes. However, in terms of the micro-F1 score, the scheme by Jochim and Schütze (2012) performed comparably well. It is an important advantage of this study that the manually annotated data corpus of the study is publicly available for future research.

### 4.2.2    Role of Linguistic Features for Classification

Some studies investigated the role of linguistic features (n-grams and cue words) in citation contexts to determine the relationship between the citing and cited publications. One of these studies was undertaken by Agarwal, Choubey, and Yu (2010), who highlighted the use of n-grams for attaining a promising accuracy. The authors annotated a corpus of 43 open-access, full-text biomedical publications, to propose an eight-category scheme. SVM and MNB classifiers were built with uni-grams (individual words) and bi-grams (two consecutive words) as their features, using the open-source Java library 'Weka.' The resulting classifier achieved an average accuracy of 92.2% and a macro-F1 of 76.5 (using n-grams). In another study from the same year, Sugiyama, Kumar, Kan, and Tripathi (2010) used the data corpus from ACL Anthology to classify sentences as either citing (i.e., including at least one citation) or non-citing. They constructed two supervised classifiers (SVM and MaxEnt) using 10-fold cross-validation with many independent features (e.g., uni-gram, bi-gram, proper nouns, previous and next sentence, position, and orthography). In contrast to the finding by Agarwal, Choubey, and Yu (2010), the authors observed that the features such as proper nouns as well as previous and next sentences are useful in classification and gave superior results, achieving an accuracy of 88.2% on the testing dataset. The results further show that the bigram feature is least useful. The limitation associated with the use of n-gram features is that they are very sparse. It might be challenging therefore to construct a model with a few overlapping features.

The importance of cue words in analysing the relationship between citing and cited publications was identified by Wang, Villavicencio, and Watanabe (2012). They used a classification approach comprising of four categories (extend, criticise, improve, and compare) – similar to that of Dong and Schäfer (2011) (see above in Section 4.2.1) – and cue phrases as features. Based on nouns, verbs, and prepositions, they considered 48 groups of cue phrases. They used a dataset of 40 articles (345 citation contexts) from *IEEE Transactions*. The results showed that more than 50% of the contexts fall into the 'extended' class following 'criticise' (30.14%), 'compare' (13.88%), and 'improve' (3.83%). Wang, Villavicencio, and Watanabe (2012) reported that a high number of cue phrases identifies the relationship between the cited and citing articles more accurately than a low number. The appearance of multiple cue phrases in the same sentence may result in low precision.

In a recent study, Small (2018) examined hedging words that best classify those citations located in either the method or the non-method sections, using a set of 1,000 biomedical articles. An LR model revealed that the frequency of hedging words such as 'may,' 'show,' and 'suggest' was higher in the citances of non-methods sections, whereas the hedging word 'using' were more frequent in the citances of methods sections. Small (2018) concluded that the predictive ability of the word 'using' in classifying citation contexts into the method and the non-method sections was higher than that of other hedging words, with a degree of accuracy of 89.5%.

The results of the studies in this section reveal that the SVM model with a specific set of features might outperform the other classifiers in determining citations accurately. Since the size of a dataset influences the features like unigram and n-gram, these features can perform better with large datasets.

### 4.2.3   Important versus Non-important Citations

In recent years, some research on classification has shifted from multiple categories to only two categories (important versus non-important). Citations that extend or use the cited work in a meaningful way are defined as important citations; citations that are used in the literature review section or compare the cited work with the citing work are defined as non-important/ incidental citations. Various authors have introduced novel classification approaches to distinguish important from non-important citations. Zhu, Turney, Lemire, and Vellino (2015) used a supervised ML approach for this task. The SVM classifier was trained with a wide variety of features, and it attained an F1 of 37% on unseen data. Though the attained F1 is not very promising, the study revealed that the feature 'total number of times a reference is cited in the citing paper' was the most important feature in identifying influential references. From the author's point of view, the selection of features is the most significant task for citation classifications. Useful features help to achieve greater accuracy and better classification performance (higher accuracy and F1).

Valenzuela, Ha, and Etzioni (2015) addressed a similar classification problem by retrieving 20,527 articles from the ACL Anthology and annotated 465 citations randomly. A range of 12 new features were extracted to train two classifiers. The model achieved an area under the curve precision recall (AUCPR) of 80% with the RF classifier. 85.4% of the citations were incidental while only 14.6% were important (influential). The authors noted that 'total per section citations' and 'self-citation' were the best predictors of an influential citation. Pride and Knoth (2017) combined the 40 features presented by Zhu, Turney, Lemire, and Vellino (2015) with the 12 features presented by Valenzuela, Ha, and Etzioni (2015) to find out the most influential features for citation function classifications. They found that a combination of just three (total number of direct citations, author overlap, and abstract similarity) led to better classification results. Similar to Valenzuela, Ha, and Etzioni (2015) and Zhu, Turney, Lemire, and Vellino (2015), they reported that the 'number of times a reference is cited in the citing paper' is the strongest predictor of a citation's influence.

**Table 7:** Features for classification (Hassan, Akram, and Haddawy, 2017: 3)

| Feature | | Description |
|---|---|---|
| | F1 | Total number of citations of a reference |
| | F2 | Number of citations in the current paper to the cited paper |
| Context-based features | F3 | Citations in introduction section |
| | F4 | Citations in literature review section |
| | F5 | Citations in method section |
| | F6 | Citations in experiment section |
| | F7 | Citations in discussion section |
| | F8 | Citations in conclusion section |
| Cue word-based features | F9 | Cue words for related work citations |
| | F10 | Cue words for comparative citations |
| | F11 | Cue words for using the existing work |
| | F12 | Cue words for extending the existing work |
| Textual features | F13 | Similarity between abstract of cited paper and text of citing paper |
| | F14 | The cited paper and citing paper share at least one author |

The annotation scheme of Valenzuela, Ha, and Etzioni (2015) provides useful guidelines for distinguishing important from non-important citations. Hassan, Akram, and Haddawy (2017) extended the feature set of Valenzuela, Ha, and Etzioni (2015) by six novel features in three categories. Table 7 depicts the complete feature set along with descriptions. The authors experimented with the five most common classifiers (NB, KNN, SVM, RF, and DT) and showed that the RF classifier with the proposed features improved the AUCPR by 4 percentage points compared to that presented by Valenzuela, Ha, and Etzioni (2015). They also evaluated each feature's importance by implementing the extra-tree classifier and realised that the newly proposed features were among the top eight features.

In a more recent study, Hassan, Imran, Iqbal, Aljohani, and Nawaz (2018) addressed the problem of citation classification by comparing two traditional ML models, SVM and RF, with the LSTM model. They applied four state-of-the-art models (Teufel, Siddharthan, and Tidhar, 2006; Abu-Jbara, Ezra, and Radev, 2013; Valenzuela, Ha, and Etzioni, 2015; Hassan, Akram, and Haddawy, 2017) and a new model for classifying citations as either important or incidental. They extracted 64 features using a dataset presented by Valenzuela, Ha, and Etzioni (2015). The results demonstrated that the proposed model improved on the state-of-the-art techniques by 11.25 percentage points. The LSTM-based deep-learning model distinguished the influential from the incidental citations with an accuracy of 92.5%.

Many studies that addressed the classification of important and non-important citations have implemented supervised machine learning techniques. Hassan, Iqbal, Imran, Aljohani, and Nawaz (2018) used another way and qualitatively clustered citations into important and non-important groups by leveraging the Self-Organizing Map (SOM). Their study is based on a dataset provided by Hassan, Imran, Iqbal, Aljohani, and Nawaz (2018). The unsupervised ML-technique SOM was deployed to obtain a qualitative understanding of the features and a good data visualisation. The SOM reduced the data to two dimensions and mapped each citation to a specific neuron (the smallest unit that performs a mathematical function). The non-important class formed an independent cluster with adjacent neurons; only 14% of the data belongs to the important class and formed independent clusters with many neurons (but not a large cluster). In terms of clustering, the results showed that it is easier to identify the non-important than the important class citations. However, the results were affected by the class imbalance issue, as only 14% of the data belong to the important class.

In a similar way to the work by Hassan, Akram, and Haddawy (2017), Tuarob et al. (2019) investigated citation context data to capture the evolution of algorithms in the scientific literature. The authors argued that in many cases, new algorithms are not developed entirely from the scratch; instead, they are built by extending the existing algorithms. Therefore, the authors classified citation contexts according to two schemes: 'utilize' (the algorithm is either used or extended in the citing document); or 'not utilize' (the algorithm is only mentioned in the citing document). A dataset of about 8,796 citation contexts was randomly selected, representing a variety of study domains, venues, and document types. To characterise the usage of the algorithms, the authors presented context-based and content-based features. The best average results (F1 = 0.905) in the binary class were achieved for the 'usage' class by the SVM classifier, which combined context and content features.

Among the studies that have addressed the problem of classifying citations into important and non-important classes using supervised classification models, the solution by Hassan, Imran, Iqbal, Aljohani, and Nawaz (2018) seems to achieve the most promising accuracy. The authors provide the extracted feature file and implementation code on GitHub for the reproducibility of their work – based on the dataset published by Valenzuela, Ha, and Etzioni (2015).

### 4.3    Citation-based Sentiment Analysis

Citation-based sentiment analysis means the classification of citations into positive, negative, and neutral classes. A citation is marked as positive if it emphasises the strength of a cited paper, and it is marked as negative if it points out the weakness of a cited paper. Neutral means that the citation is rather descriptive in nature.

### 4.3.1    Context Window Selection for Sentiment Classification

Several studies have investigated the influence of the context window size on the performance of automatic systems for classifying citation sentiments. Athar and Teufel (2012a) observed the importance of context while analysing the sentiments of citations, using a dataset of 1,741 citations from ACL Anthology. They suggested two methods for context utilisation. First, they considered the citing sentence plus a context window (one sentence before the citation, citing sentence, and the two sentences after the citation) to train a classifier on the merged texts after extracting features (1 to 3 n-grams). Second, each sentence in a context window of four sentences was treated as an individual sentence and was annotated using a four-class (positive, negative, neutral, and exclude) annotation scheme. The study of the authors revealed that sentiment analysis achieves better results when it does not consider the merged context. The context window of four sentences, which was treated as an individual sentence, provided useful information for sentiment detection.

In contrast to this first study, Athar and Teufel (2012b) automatically identified the sentiments of cited papers using n-gram and dependency features on an annotated benchmark corpus by Athar (2011). They compared their approach of a four-sentence context to that of a single sentence and proved that overlooking citation context would affect the accuracy of a sentiment system, especially with respect to citations including criticism. They found that a system based on a single sentence results in a loss of sentiment, due to lesser available information. The results generally point out that a system based on four-sentences contexts should be favoured in sentiment classifications.

### 4.3.2   Role of Linguistics features for Sentiment Classification

While some studies have focused on the influence of the context window size, others have investigated the usefulness of (linguistic) features for citation sentiment analysis. Athar (2011) examined the effectiveness of novel features, including n-grams, dependency relation, scientific lexicon, and sentence splitting (by utilising citing sentences only). Of the 8,736 annotated citations, 1,472 were used as a training corpus and the rest as a testing dataset consisting of 244 negative, 743 positive, and 6,277 neutral citations. The author noted that tri-grams and dependency features are best at automatically identifying a citation's sentiment, achieving a macro-F1 of 89%.

Abu-Jbara, Ezra, and Radev (2013) annotated the ACL dataset to train the SVM classifier with 10-fold cross-validation techniques and a list of features (see Table 8). The authors undertook an analysis of each feature's importance using the chi-squared test and concluded that features associated with subjectivity, such as negation, subjectivity cues, and speculation, are important for the classification of polarity achieving an F1 value of 74%. While the polarity features proposed in the study by Abu-Jbara, Ezra, and Radev (2013) did not outperform the model proposed by Athar (2011), one can conclude from the results that the consideration of context improves the model's accuracy significantly.

**Table 8:** Features used for analysing citation purposes and polarity (Abu-Jbara, Ezra, and Radev, 2013: 601)

| Feature | Description |
| --- | --- |
| Reference count | Number of references that appear in the citation context |
| Is separate | Whether the target reference appears within a group of references or separate (i.e., single reference) |
| Closest verb/ adjective/adverb | The lemmatized form of the closest verb/adjective/adverb to the target reference or its representative or any mention of it. Distance is measured based on the shortest path in the dependency tree |
| Self-citation | Whether the citation from the source paper to the target reference is a self-citation |
| Contains 1st/3rd person pronoun | Whether the citation context contains a first/third-person pronoun |
| Negation | Whether the citation context contains a negation cue |
| Speculation | Whether the citation context contains a speculation cue |
| Closest subjectivity cue | The closest subjectivity cue to the target reference or its representative or any anaphoric mention of it |
| Contrary expressions | Whether the citation context contains a contrary expression |
| Section | The heading of the section in which the citation appears |
| Dependency relations | All the dependency relations that appear in the citation context |

As the annotated dataset compiled by Athar (2011) achieved promising accuracy, this dataset has been used by many studies as a gold-standard to analyse the polarity of citations. One of these studies has been published by Ikram, Afzal, and Butt (2018). They proposed a classification model and explored the usefulness of n-grams in citation-based sentiment analysis. Their results showed that a higher value of n-grams (n = 5) yielded 2 percentage points better scores in determining the sentiment of citation contexts than lower values. The authors compared the efficiency of the proposed model with other available commercial tools for citation polarity, i.e., SEMANTRIA (a Microsoft Excel add-in) and THEYSAY (an online sentiment analysis tool). Their results revealed that the model followed the same trend line as depicted by SEMANTRIA and outperformed the results of THEYSAY by achieving an F1 score of 85.91%. Although SEMANTRIA produced the most precise predictions (F1 score of 96%) on the dataset by Athar (2011), it produced very different results across the datasets. Thus, it seems that SEMANTRIA has some weaknesses.

**\*Corresponding author's email:** saeed-ul-hassan@itu.edu.pk

Ikram and Afzal (2019) used the dataset from the field of computer science presented by Athar (2011) and another from the bioinformatics domain. In the dataset from the bioinformatics domain, 285 papers were randomly selected containing 3172 neutral, 702 positive, and 308 negative citations. Ikram and Afzal (2019) extracted different POS (nouns, proper nouns singular, proper nouns, determiners verbs, and adjectives) from citing sentences to analyse their sentiments. Various ML classifiers were trained in aspect-based sentiment classification with n-gram features. The results showed that the SVM classifier outperformed the other classifiers, with a precision and recall of 85.6% and 85.2%, respectively. Similar to Ikram, Afzal, and Butt (2018), they suggested applying a high value of n-grams (about n = 5) to achieve the best results.

### 4.3.3 Influential Features for Sentiment Classification

Using a range of various features (e.g., closest verb or self-citation) and machine learning classifiers, some studies have employed sentiment classification models (Mäntylä et al., 2018; Pang and Lee, 2008). For example, Teufel, Siddharthan, and Tidhar (2006) designed an annotation scheme for the automatic classification of citation polarity, using cue words as a feature. They conducted two classification schemes: a four-way scheme (weak, contrast, positive, and neutral) and a three-way scheme (positive, negative, and neutral). The dataset was comprised of 116 conference papers from the Computation and Language e-print archive. The study revealed that 83% accuracy was achieved by the three-way classification, compared to 75% by the four-way classification. Hence, the three-way classification scheme produced more constructive results.

Piao, Ananiadou, Tsuruoka, Sasaki, and McNaught (2007) suggested the polarity relation between cited and citing publications (i.e., attitudes of authors' approval or disapproval for the work they cite) is useful for information retrieval and text mining tasks. They designed a system for authors to search for publications in an extensive collection of articles. For this purpose, the authors collected and mapped citations (subjective words and sentiments) to form a network combined with an opinion polarity relation by employing semantic lexicon resources and NLP tools. The proposed system gathered cited articles and compiled a citation distribution list that shows the opinion polarity relations between articles and citations. Although the system appears promising, a drawback of this study is that the authors neither mentioned the accuracy of their model nor compared their model with other baselines.

### 4.3.4   Class Unbalancing in Sentiment Classification

The results by Li, He, Meyers, and Grishman (2013) highlight the class unbalancing issue in sentiment classification of citations and its effects on the performance of the classification models. They suggested a unique scheme that includes a total of eight positive, one negative, and three neutral functions to study the sentiment of citations in PubMed. To classify a citation's function automatically, they used a ME-based approach with various syntactic and surface features. The model achieved a low accuracy of 67%. They concluded that its poor performance was due to the class imbalance in the data corpus. To solve the class imbalance problem, Sula and Miller (2014) eliminated the neutral class from the scheme (to have only meaningful classes) and build a system to measure the positive and negative relationships between citing and cited publications. They used a data corpus of 159 documents from four prominent humanities journals: *Art Bulletin*, *Language*, *Journal of Philosophy*, and PMLA (*Journal of the Modern Language Association of America*). The NB classifier was trained on two datasets (the positive set contains 176 citations and the negative set 58 citations) along with features such as frequency and in-document locations to examine the polarity of citations. The study revealed that negative and positive training sets should be larger to increase the accuracy of classifiers.

To solve the class imbalance problem, Jha et al. (2017) also decided to eliminate the neutral class, since more than half of the citations were in this class. Using the ACL dataset, they set up two binary classification schemes with a battery of features inspired by the work of Abu-Jbara, Ezra, and Radev (2013). Citations were classified as subjective (polarized) or objective (neutral); subjective (polarized) citations were categorised as positive or negative. The results showed that the polarity classifier yielded more intuitive results than the subjectivity classifier. As the dataset was highly skewed (more than half of the citations were neutral), eliminating the neutral class increased the accuracy by up to 6 percentage points. The chi-squared analysis led to a similar conclusion as by Abu-Jbara, Ezra, and Radev (2013): subjectivity features, such as subjectivity cues, negation, and speculation, are more important than other features.

Taşkın and Al (2018) created an automatic classification scheme for Turkish citations. To classify the citation context, NB and RF were tested using a 10-fold cross-validation, and various semantic and syntactic features. The results showed that the RF classifier accurately detected 96% of the positive and 70% of the negative citations, with an average accuracy of 89%. As the training dataset contains a lesser number of negative instances (only 0.8%) – similar to Sula and Miller (2014) – they believed that increasing the number in the negative class training dataset would increase the classifier's accuracy.

The results of the studies in this section point out that removing the neutral class would lead to a better accuracy of classification schemes. However, the classifier should have enough data to train. Especially the training sets for the positive and negative classes should be large and balanced in number.

## 4.4 Citation-based Summarisation

Automatic text summarisation might be a way of producing a fluent and concise summary of a publication by capturing the overall meaning (and critical content). The summarisation approaches should be able to identify both the uniqueness and similarity between publications. A typical study in this area has been published by Verma and Lee (2017). The authors compared human-generated with automatically-generated summaries. Their results revealed that, on average, 9% of the words in human-generated summaries do not appear in the original article. Thus, it seems that human-generated summaries have added values compared to automatically-generated summaries.

CL-SciSumm (the first medium-scale shared task on scientific document summarisation in the computational linguistics domain) divides the task of automatic summarisation into three categories: (1) search of relevant text-spans that identify the relationship between citing and cited document, (2) classification of discourse facets, and (3) generating of abstractive summaries (Chandrasekaran et al., 2019). We discuss each task in the following subsections.

### 4.4.1 Baseline Summarisation Models

Various summarisation techniques have been used hitherto as a baseline for summarisation tasks. The MEAD summarisation system (Radev et al., 2004) – the most elaborate publicly available platform for multi-lingual summarisation and evaluation – is a centroid-based summarisation system. The MEAD summariser has three components: first, feature extraction and its conversion into feature vector; second, conversion of feature vector into scalar value, and third, assignment of scores to sentences, and addition of similar sentences to the summary. Another technique named as LexRank summarisation (Erkan and Radev, 2004) builds a graph of all candidate sentences and then evaluates the meaningful sentences using eigenvector centrality. Although the edge between two candidate sentences might demonstrate the similarity between the sentences (and thus their meaningfulness), the authors proposed the LexRank system as sensitive to noisy data.

Two variants of the LexRank system have been proposed by Qazvinian and Radev (2008) for the summarisation of publications: Cluster Lexrank (C-lexrank) calculates LexRank within each cluster while Cluster Round-Robin (C-RR) picks sentences from each cluster in a Round-Robin way. The Round-Robin way starts with the largest cluster; then, the sentences are extracted in the order they are listed in each cluster. In the C-lexrank model, first, the reliable information (the 'nuggets') from each paper are prioritised manually, and second, a weight is assigned to each 'nugget' based on the evaluation. The results of the study show that the C-lexrank model attains an increased 'nuggets' based-pyramid score compared to the C-RR and LexRank model (6 and 4 percentage points, respectively).

Earlier this decade, Abu-Jbara and Radev (2011) presented an approach in which they divided the task of automatic document summarisation into three steps. The first step included the tasks of reference tagging, context identification, and sentence filtering. In the second step, the extracted representative sentences were classified, similar sentences were added into a cluster, and the LaxRank value of each sentence was computed. In the third step, the sentences were added into a summary based on the sentence ranking of cluster and LaxRank values. The authors used a dataset of 55 papers from the ACL Anthology Network and deployed a well-known evaluation metric called Recall-Oriented Understudy for Gisting Evaluation (ROUGE). They compared the proposed technique with the techniques used by Radev et al. (2004), Erkan and Radev, (2004) and Qazvinian and Radev (2008). The results demonstrated that the approach of the authors achieved a 12, 13, and 10 percentage points improved ROUGE score, respectively.

### 4.4.2   Importance of Citation Text for Summarisation

Citations are considered to be a significant source of information for generating automatic summaries in various text-mining areas (Li et al., 2017; Teufel and Moens, 2002). Teufel (2006) argues that citations include valuable subjective assessments of cited publications. These assessments can be exploited to generate a summary. Other authors agreed with Teufel (2006) and investigated the effectiveness of citing sentences in the task of summary generation such as Elkiss et al. (2008). These authors used the cosine similarity metric and examined the usefulness of citing sentences and abstracts in an automatic summary generation. Their study is based on the PubMed Central dataset of 2,497 articles. The authors concluded that – in the absence of an abstract – citing sentences may be a good substitute for automatic summarisation.

Using the ACL dataset, Mohammad et al. (2009) analysed the importance of citing sentences in summarisation by comparing it with abstract-based summaries and full-text summaries. They used two different approaches for evaluating automatic summarisation: the first approach is ROUGE, and the second approach is a nugget-based pyramid evaluation. In this evaluation, the elements of valuable information, the so-called 'nuggets,' from each article were prioritised manually. Then, a weight was assigned to each 'nugget' based on this evaluation. The ROUGE evaluation results showed that summaries generated from abstracts performed significantly better ($p < 0.05$) than summaries generated from citation contexts. Moreover, summaries generated using citation contexts obtained significantly better scores ($p < 0.05$) than those generated from full-text publications. The results from the nugget-based pyramid evaluation demonstrated thus that the summaries generated from citation contexts outperform those generated from abstracts and full-text publications. Mohammad et al. (2009) concluded that both abstracts and citation contexts have unique information that can be used to improve summarisation results.

The study by Qazvinian et al. (2013) illustrated the importance of citing sentences for the creation of summaries for Question Answering (QA) and Dependency Parsing (DP) articles. Using state-of-the-art techniques, as applied by Mohammad et al. (2009), the study generated summaries based on three sets of information: abstract only, citation only, and full paper. For the evaluation, Qazvinian et al. (2013) used the pyramid score 'ratio of the sum of the weights of semantically relevant words to the sum of the weights of an optimal summary'. The study demonstrates that the generation of technical summaries benefits considerably from the use of citing sentences.

The studies reviewed in this section reveal that there is a small but quantifiable difference in the information content provided by citation contexts as compared to abstracts or full-texts. The results of the studies indicate therefore that the use of citing sentences might lead to improved extractive summaries of publications.

### 4.4.3 Identification of Text-Span (Task 1A)

We mentioned above the merits behind the extraction of citations in the running texts. Numerous studies addressed the task 1A to generate a summary of contributing publications using the dataset from CL-SciSum, which comprised 40 annotated sets of citing and cited papers from the CL domain.

Kaplan, Iida, and Tokunaga (2009) presented a co-reference, chain-based approach with cosine similarity as a feature to extract the 'citation-site (c-site),' which is the block of text that contains both the citation and its context. The authors compared their approach with two baseline models (baseline 1 extracts only the citing sentence; baseline 2 extracts the sentences before and after the citing sentence) using a labelled dataset containing 38 articles from the ACL dataset citing four publications. The results demonstrated that the proposed approach achieved a high micro-average F1 score of 84%. However, it should be considered in the interpretation of this result that the study is based on a small and less representative dataset containing only 94 citing sentences. Qazvinian and Radev (2010) addressed a similar problem of contextual information extraction from publications as Kaplan, Iida, and Tokunaga (2009). The authors suggested a framework based on probabilistic inference. They modelled lexical similarities of sentences as a Markov random field to discover the patterns created by the context data. They employed a belief propagation mechanism to identify those sentences with the same context. For experimentation, they used ten papers from the ACL anthology. Their results demonstrated that the use of citation contexts (four sentences on each side of the reference anchor) with citing sentences improved the pyramid score considerably, from 0.41 to 0.63. Although the studies by Kaplan, Iida, and Tokunaga (2009) and Qazvinian and Radev (2010) used rather small datasets, they reveal that the generation of fluent scientific summaries is a non-trivial task in the absence of sufficient background information.

The study by Nomoto (2016) worked on the extraction of relevant citation texts using two hybrid models: a Term Frequency-Inverse Document Frequency-based (TF-IDF, a weighting scheme used in text mining to evaluate the importance of a word for a document in a corpus) similarity model; and a single-layer ANN that scores relevant citing text more highly than irrelevant text. The results showed that the performance of the ANN-based model is better than that of the TF-IDF-based model. Nomoto (2016) suspected that the TF-IDF model's performance is lacking due to its inability to define false and true targets clearly. There may be some words that appear in both true and false targets, which could quickly derail the accuracy of classifiers. In a similar study to that by Nomoto (2016), Klampfl, Rexha, and Kern (2016) compared three different approaches for classification: (1) modified-tsr, where the TextSentenceRank algorithm is applied to the data. This algorithm is a modified version of the TextRank algorithm, which is a graph-based ranking algorithm to extract key sentences or key terms. (2) tsr-sent-class, where a binary classifier is used to decide for each candidate sentence, whether it can be served as a reference text span. (3) sect-class-tsr, where a binary classifier is used to decide for each section, whether it contains a reference text span. The authors found that modified-tsr achieved the best results by extracting both the 'most relevant' key terms and sentences.

In this section, we summarised some studies addressing the task 1A of automatic summarisation. The comparison of the different approaches reveals that the TextRank algorithm approach presented by Klampfl, Rexha, and Kern (2016) seems to be a promising method for the extraction of text spans.

### 4.4.4 Facet Identification (Task 1B)

Addressing the task 1B of CL-SciSumm, some studies have considered the facet identification problem (i.e., the classification of referenced text spans into the following classes: 'implication,' 'method,' 'aim,' 'results,' and 'hypothesis') as a text-classification problem. The studies have used various modelling approaches to solve the facet identification problem based on the CL-SciSumm 2016 dataset.

The facet identification problem was targeted by Cao, Li, and Wu (2016), who stated that this problem is a multi-label classification task. They noted that the facet distribution of the dataset is exceptionally imbalanced, as 60% of the text spans are in the methods section and only 9% in the hypothesis section. They stated that DT could remember the patterns of all facets, whereas SVM and NB remember only those of the dominant class. Therefore, the DT classifier was employed (featuring TF-IDF vectors) and achieved a low micro-averaged accuracy of 59%. The study by Lu, Mao, Li, and Xu (2016) presented a similar issue with class unbalancing. The authors proposed a feature engineering approach to text-span classification and defined a set of features, including lexical features such as TF-IDF, topic similarity (cosine similarity between the citation and reference span), concept similarity, and sentence importance. After the extraction of the citation context, the context was classified based on the section of the article or the 'discourse facet' in which the text appears by applying the DT, SVM, and NB classifiers. In contrast to the study by Cao, Li, and Wu (2016), the authors revealed that the NB classifier achieved a balanced performance across all five facets while the SVM classifier achieved the highest micro-average accuracy of 65%.

Cao, Li, and Wu (2016) and Ma, Xu, and Zhang (2018) proposed multi-stage actions for facet classification. First, they extracted similarity-based, position-based, and rule-based features for multi-label classifications. Second, they applied a sampling-based algorithm to pre-process the imbalanced data corpus. Third, each facet built a dictionary and assigned the reference span to the facet whose dictionary contained any span words. The results showed that the proposed system achieved an average precision of 0.7169 – a better result than that achieved by Cao, Li, and Wu (2016) and Lu, Mao, Li, and Xu (2016) – in identifying facets for summarisation using an imbalanced dataset. Overall, the study concluded that similarity-based features are more suitable than position-based features for summarisation. Moreover, in working with the class imbalance dataset, TF-IDF similarity and IDF similarity turned out to be two robust features.

### 4.4.5 Generating Summaries (Task 2)

Some studies in our publication set attempted to solve the task 2. Mei and Zhai (2008) worked on the problem of summarisation and proposed a Language Modelling method (LM: scoring matches between the queries and documents) to extract those sentences from a publication that represented the most critical content. The study was undertaken with a relatively small dataset of only 14 articles from the Medical Literature Analysis and Retrieval System Online (MEDLINE) database. The small dataset is an explicit limitation of the study. The authors formed a 'citation document' by concatenating all citation contexts in an article. To generate a summary from the original article and all its citation documents, they used a multi-document summariser. Their results showed that the language-based model performs better than the conventional summarisation techniques (MEAD).

A similar LM approach for the classification of citation contexts as used by Mei and Zhai (2008) was proposed by Tandon and Jain (2012). Their LM model was trained on 30 articles from the MA search engine. The model used an opinion vocabulary with two types of terms (uni-grams and bi-grams) to indicate the context and the adjectives, verbs, and adverbs that describe the cited publication's opinion. In contrast to the findings of Mei and Zhai (2008), the study showed that a combination of adjectives, verbs, and bi-grams models beats the accuracy of the LM model, achieving a 68.54% average precision. The proposed LM model attained an average precision of 67%.

Based on the TextRank algorithm, further improvements for summarisation have been proposed by Barrera and Verma (2012). They suggested a combination of syntactic, semantic, and statistical methods using a dataset from the Document Understanding Conference (DUC) 2002 and scientific magazine articles. The study is based on three-position models: For the first model, they used sentences close to the beginning and end of an article. For the second model, they prioritised the use of the first section of an article. For the third model, sentences close to headings were taken. The authors concluded based on their empirical results that semantic linkage and topic-heading relevance produce useful summaries. The third position model exceeded the performance of TextRank and MEAD with an F1 measure of 0.71.

Conroy and Davis (2015) compared the performance of a vector space model and their proposed Non-Negative Matrix Factorization model in the study with baseline language modelling approaches. The vector-based model used the Term Frequency model (a simple vector space model for text) and Non-Negative Matrix Factorization (an algebraic dimensionality reduction technique) to estimate the weights of terms for automatic document summarisation. The results showed that the Non-Negative Matrix Factorization rank improved the ROUGE scores from 0.065 to 0.074 as compared to the Term Frequency model.

The recent work by Yasunaga et al. (2019) provided the first large-scale manually-annotated corpus for paper summarisation. The corpus consisted of 1,000 sample articles, including citation information and human summaries of 150 words. The authors proposed hybrid models that integrated both the research community's views (citations) and authors' original highlights (abstracts). The results of the study demonstrated that the hybrid models generated more comprehensive summaries than traditional citation-based and abstract-based summaries, achieving a recall of 41.69%.

### 4.5 Citation Recommendation System

For the task of recommending citations, citation recommendation systems match user queries with existing publications in a database and recommend publications that could be cited by the user. According to the different user inputs, these systems can be divided into three main types: keywords-based citation recommendation, citation-list-based citation recommendation, and context-based citation recommendation. Keeping the scope of this review in mind, we focus only on context-based citation recommendation systems in this section. These systems take a few sentences as input and recommend a list of articles as possible citations based on the local context of input sentences. Context-aware approaches estimate the semantic similarity between an article and the citation contexts of articles citing that document. Articles to cite are recommended then, based on the similarity score.

Studies that have addressed the development of citation recommendation systems have used two techniques: topic modelling and neural network. Explanations of these two techniques are provided in the subsequent sections.

### 4.5.1 Topic Modelling

Early research on citation recommendation systems formalised the task of finding relevant publications to cite as topic discovery. Nallapati, Ahmed, Xing, and Cohen (2008) addressed the problem of listing documents in recommendation systems as topic modelling. They proposed the Link-PLSA-LDA model that is a combination of Latent Dirichlet Allocation (LDA: a statistical model to assign the text in a document to a particular topic) and Probabilistic Latent Semantic Analysis (PLSA: a statistical technique for analysing two-mode and co-occurrence data). They experimented with this model on a CiteSeer dataset including 3,312 documents with a vocabulary of 3,703 unique words. Their results showed that the proposed model performs better than the traditional topic modelling techniques (i.e., LDA and PLSA). However, the model was computationally expensive as it considered the links between every document.

Tang and Zhang (2009) also targeted the task of ranking the relevance of papers in a recommendation system for a given citation context. They proposed a two-layer restricted Boltzmann machine model known as RBM-CS, which discovers the relationship between articles based on the topics of the articles given as input. The results demonstrated that RBM-CS achieved 0.4237 MAP and performed better than LM-based techniques for citation recommendation. The desired accuracy of the model could not be attained, however, due to data sparsity issues (i.e., certain expected values were missing in the dataset).

The approach by He, Pei, Kifer, Mitra, and Giles (2010) generated a candidate citation set by taking co-authorship relations into account. The input of the system is a query document, and the system automatically discovers the positions at which citations should be located and recommends the documents for citing with high similarity scores. Using the CiteSeerX dataset, the system outperformed the baseline methods (uni-grams, bi-grams, dependency model, etc.) with an average precision of 47.2%. The authors claimed that other existing approaches do not perform well due to various noise (irrelevant words) in documents.

For developing ambitious citation recommendation systems based on topic modelling techniques, the data sparsity and noise issues ought to be resolved. Wang and Blei (2011) presented such a system: a collaborative topic regression model that combines the merits of probabilistic topic modelling and traditional collaborative filtering. They investigated the model using a relatively small dataset including 37 articles. The empirical results were promising: the proposed model was able to make relatively useful recommendations.

### 4.5.2 Neural Networks

In recent years, several studies have started to deploy neural networks for citation recommendations. In these systems, the neural networks have been trained on the contextual content of papers or by linkages in bibliographic networks. In one of the first studies, Huang et al. (2015) proposed the multilayer neural network model that learns document representations and words to compute the probability of citing a document for a given citation context. For implementing the model, the CiteSeer dataset containing 10,760,318 citation contexts was used. The results based on this dataset revealed that the model could improve the overall recommendations with a 9 percentage points gain on recall compared with the LDA model.

Gupta and Varma (2017) addressed the problem of article recommendation by introducing a novel approach that combines article content with a graph structure. They used a dataset from Arnetminer (a database that contains research articles published between 1958 and 2014) to test their approach. The results demonstrated that the approach outperforms state-of-the-art techniques (i.e., TF-IDF and LDA) by 21 and 39 percentage points, respectively, in terms of MAP.

In the most recent study, Yang et al. (2018) used an LSTM model to improve citation recommendation systems. The authors proposed a model that, first, learns the embedding of citation contexts, second, measures their relevance, and third, selects documents of the highest relevance. Experiments were conducted using the ACL Anthology Network and Digital Bibliography and Library Project (DBLP) datasets. The results revealed an increase of 2 percentage points compared to the effectiveness of the topic-modelling technique.

The comparison of the employed neural networks in the various studies (presented in this section) concerning their ability to improve citation recommendation systems is difficult: since each study has used another dataset with different parameters, the results are scarcely comparable. However, it seems that neural network techniques are promising solutions that might outperform topic modelling-based techniques (i.e., LDA).

## 5 Discussion

Citation analysis is one of the most crucial methods in research evaluation. This is reflected, for instance, in the utilisation of citation data in several international university ranking systems. The use of citations in research evaluation is mainly based on the normative theory of citing. According to Merton (1973), citations can be interpreted as a reward for intellectual achievement. However, several studies have shown that many citation decisions do not seem to follow this norm. Authors have various reasons to cite, and citations can have many functions in the scholarly communication process. For instance, authors may cite reputable researchers to give more weight to their own results and ideas.

Various forms of context and content analyses have been used to investigate the reasons for citing and functions of citations. Early studies speculated about possible reasons and functions and undertook first empirical investigations into citation context and content based on small datasets. In recent years, various advanced techniques (NLP and ML techniques) have been developed for citation context and content analyses which have been used in different contexts of specific applications. This review focuses on these developments and provides a survey of the various mining techniques for citation content/context identification and classification, sentiment analysis of citations, as well as experimentation and evaluation of related search areas such as citation-based recommendation systems and citation-based summarisation. Scholars from various fields such as life science, medicine, and engineering have used the NLP and ML techniques for these analyses and applications. One important reason for the increasing popularity of the techniques is the availability of full-text articles in an increasing number of data repositories such as ACL Anthology and PLoS (that can be freely used for research projects).

The studies that we have included in this literature overview addressed several topics. In this section, we summarise the most important findings for these topics and discuss their (practical) implications. We will begin the summarisation with the challenges that are associated with citation in-text analyses. First, a citation context might contain multiple references to publications, and only a part of a specific context might be relevant for the focal publication. In these cases, it is often a difficult task even for human annotators (domain experts) to assign the context to a specific class. Second, most of the publications in this research area used data from the ACL Anthology. There is a lack of variance in the employed datasets, therefore, raising questions concerning the generalisability of the findings. Third, while some databases are available with full-text access today, there is limited access to several full-text datasets due to copyright restrictions.

Several studies have investigated the optimal length of citation contexts for analysing the relationship between citing and cited articles. The results have shown that extended citation contexts appear to be more effective than shorter citation contexts (Aljaber et al., 2011). The extended citation contexts contain more descriptive terms from the citing article (Ritchie, Robertson, and Teufel, 2008). It seems that the optimum window size for the context analysis is four sentences. Alongside ascertaining the optimal length of citation contexts, many authors have also analysed the usefulness of various lexical and structural features in context extraction. In terms of the accuracy attained by the features, it seems that lexical features (determiners and conjunction adverbs) are more significant than structural features (position and reference) for context extraction (Abu-Jbara, Ezra, and Radev, 2013). Numerous studies have used ML techniques for citation classifications with preferred features, such as linguistic features, cue words, contextual features (closest noun phrase and conjunctive adverb), and location features (position and section). The outcomes point out that linguistic features and cue words are effective in achieving high evaluation scores.

In this literature overview, we have reviewed studies that applied NLP and ML techniques to analyse citation sentiments such as lexicon-based methods (e.g., SentiWordNet). These methods are used to search for terms that reveal the sentiments and classify citation contexts into different polarity levels. The studies have utilised various feature-sets: the uni-gram feature seems to be very popular whereby higher values of n-grams lead to better classification results (Ikram, Afzal, and Butt, 2018). In the studies reviewed here, SVM and NB classifiers are the most frequently used models for sentiment classifications (Abu-Jbara et al., 2013; Athar, 2011). Their classification results might be improved by eliminating the neutral class in the analyses. Most of the studies have reported that class imbalances' effects lead to a decrease in the accuracy of the employed classifiers. Another problem which also decreases the classifiers' accuracy is that the studies are based on small datasets (to reduce the annotation efforts). Against the backdrop of these problems, we believe that the use of large publication sets for citation sentiment tasks is an important future research direction by leveraging deep learning methods.

Citing sentences can contain additional information about the scientific document that has been cited. Therefore, the information can be used to characterise cited documents in addition to the information extracted from abstracts. In studies on citation-based summarisation which we included in this overview, a considerable weightage has been given to the citing sentence in summarisation as compared to solely abstract-based or full-text summaries. Since these studies have shown that the information content of citation contexts is lower than that of abstracts (or full-texts), citation contexts should be preferred to be used in conjunction with other information (Qazvinian and Radev, 2010). In the absence of an abstract for summarisation, however, citation context may serve as a substitute (Elkiss et al., 2008).

The studies that have dealt with automatic summary generation in the past (as an alternative to a human-generated summary) classified this task into three steps: reference-span identification, reference-span classification, and summary generation. For reference-span identification, the TextRank method may be preferred since it outperformed other state-of-the-art approaches, as discussed by Klampfl, Rexha, and Kern (2016). For the task of facet identification, studies have revealed that the distribution of facet is highly imbalanced. The Naïve Bayes classifier achieved a low accuracy, but for a balanced performance among all facet, the SVM classifier achieved a high accuracy (Lu, Mao, Li, and Xu, 2016). The studies dealing with summary generation have shown that the language models are more beneficial than the baseline models such as MEAD and LexRank. The authors of these studies took into account linguistic patterns to generate summaries that are not just a combination of sentences (Ma, Xu, and Zhang, 2018).

There are several limitations associated with automatic summarisations: most of the studies have utilised relatively small datasets to tackle the summarisation problem. The use of large publication sets is vital for future research direction, which might lead to better generalisability and applicability of the results. We assume – based on the reviewed studies – that neural abstractive models need to be increasingly exploited to generate automatic summaries of publications by using deep-learning models (See et al., 2017). The increasing availability of full-text data and the rapid development of ML and deep-learning methods bring new challenges in citation sentiment analyses, such as multi-lingual summarisation (the processing of several languages and providing of summaries in a single language as output) and multi-document summarisation (the processing of several documents to compile single summary documents as an output). There has been a surge of novel methods for automatic summary generation as well as summary evaluation methods such as Rouge-1 (Lin and Hovy, 2003) or BLUE Score (Dreyer and Marcu, 2012). However, there is a need for evaluation metrics that can determine whether the generated summaries are coherent, sentence-wise, and grammatically correct.

Search engines and database providers utilise citation contexts to show how publications are related to each other. One goal of these analyses is to find relevant publications for citation recommendation systems. Our overview of the literature in this area shows that the studies mainly utilise two techniques to predict citations: topic modelling and neural networks. Topic modelling techniques such as LDA are the traditional ways of recommending citations. The problem with these techniques is yet that they do not perform well based on large and noisy datasets (He, Pei, Kifer, Mitra, and Giles, 2010). Therefore, to recommend citations based on larger datasets, neural networks (RNN, LSTM) have been used – trained on the contextual contents of papers or by linkages of papers in bibliographic networks. The empirical comparison of neural networks techniques with topic modeling-based techniques (i.e., LDA) has shown that neural networks-based recommendation systems have an at least 2 percentage points better MAP than topic modeling-based recommendation systems.

One of the main challenges of research on citation recommendation systems is still to reliably and validly capture the meaning of citation contexts and to find the relevance of the cited document for the author of the citing document. Advanced NLP (Batista-Navarro, Kontonatsios, Mihăilă et al., 2013; Thompson, Nawaz, McNaught et al., 2017) and Deep Learning (Jahangir, Afzal, Ahmed et al; 2017) techniques might be helpful in this context (Ananiadou, Thompson and Nawaz, 2013; Shardlow, Batista-Navarro, Thompson et al., 2018).

## 5.1   Concluding Remarks

This literature overview could reveal that the applications of NLP and ML techniques are significant and important developments in the area of citation context and content analyses. However, the studies undertaking these analyses were constrained by several limitations. The first and foremost limitation is the tedious annotation task of the texts. While the manual tagging for training the classifier is a lengthy and challenging process, the automatic tagging produces poor results. Second, in many studies, scholars from various disciplines (such as social sciences, engineering, or health sciences) annotated field-specific datasets and employed NLP and ML models in their respective fields. Thus, cross-model comparisons of the various approaches are limited. Interdisciplinary research groups could be formed to annotate (comprehensive) standard datasets including papers from many fields. Third, only a few studies have published their code and dataset hitherto, making it virtually impossible to compare and replicate the empirical findings. Thus, research in this area could be more openly done. Fourth, one of the most critical challenges in automatic citation content/context studies is the difficulty of extracting the citation context from the publications due to the variability in citation styles (which might differ between journals, publishers, and fields). The algorithms used for automatic approaches to obtaining citations may miss those citations that fail to adhere to the searched style. Thus, approaches should be developed in future research that tackles this problem.

## 5.2   Guidelines for the application of NLP and ML techniques

In the final section of this review on studies that have used NLP and ML techniques for citation context and content analyses, we provide a step-by-step guideline on how these techniques can be used in practice:

The first task refers to the process of data collection: for citation context/content analyses, it is required to analyse full-text publications. Nowadays, some data repositories provide such full-text publications that can be used for these analyses. For example, the ACL Anthology includes more than 23,000 full-text publications with more than 350 venues. Another accessible dataset is provided by the open-access publisher PLoS. Currently, this dataset includes more than two million publications ranging across 200 disciplines.

The second task refers to the extraction of the citation context information, which is well manageable with standardised datasets such as those from ACL Anthology or PLoS. However, it is a difficult and challenging task with non-standardised datasets, primarily due to the possible use of multiple citation formats and styles. So far, no off-the-shelf solution can be used to sufficiently tackle this problem in the process of citation context extraction. An essential question in this process concerns the size of the context data around the citation anchor. We recommend (based on the results of the corresponding studies) to apply the four-sentences approach proposed by Athar and Teufel (2012): the citing sentence, one sentence before the citing sentence, and two sentences following the citing sentence should be extracted. The citation anchor can be identified by using regular expressions; the surrounding sentences can be extracted as desired then.

The third task concerns the annotation of the dataset for training the classifiers. We recommend using the annotated dataset published by Jochim and Schütze (2012) for citation classification, the classification of Valenzuela, Ha, and Etzioni (2015) for identifying important and non-important citations and the baseline by Yasunaga et al. (2019) for citation-based summarisation.

The fourth task refers to feature computation, which is a vital but time-consuming task. Many authors have not provided their datasets and codes which could be used as starting points for subsequent studies. Since Hassan, Imran, Iqbal, Aljohani, and Nawaz (2018) are one of the few authors who have published their dataset and code (on the GitHub repository), we recommend drawing on that material.

The fifth task concerns the necessity to make codes and datasets available. In order to stimulate research activities based on NLP and ML approaches, we would like to encourage all authors to make their annotated datasets and codes freely available.

## Acknowledgment

The authors (Salem Alelyani and Saeed-Ul Hassan) are grateful for the financial support received from King Khalid University for this research Under Grant No. R.G.P2/100/41.